# Tunable interplay of orbital and spin magnetization in trigonal tellurium

Zhenqi Hua[1, 2, *], Chang Niu[3, 4, *], Sandeep Joy[1, 2, *], Pukun Tan[3, 4], Gang Shi[1], Haoyang Liu[1], Jiaxing Guo[1], David Graf[2], Peide Ye[3, 4], Cyprian Lewandowski[1, 2, †], and Peng Xiong[1, †]

1. *Department of Physics, Florida State University, Tallahassee, FL 32306*
2. *National High Magnetic Field Laboratory, Tallahassee, FL 32310*
3. *Elmore Family School of Electrical and Computer Engineering, Purdue University, West Lafayette, Indiana 47907*
4. *Birck Nanotechnology Center, Purdue University, West Lafayette, Indiana 47907*

* Z.H, C.N., and S.J. contributed equally to this work

† clewandowski@magnet.fsu.edu, pxiong@fsu.edu

**Abstract**

Orbital effects, despite their fundamental significance and potential to engender novel physical phenomena and enable new applications, have long been underexplored compared to their spin counterparts. Recently, surging interest in the orbital degree of freedom has led to the discovery of a plethora of orbital-related effects, underscoring the need for a deeper understanding of their roles in quantum materials. Here, we report systematic experimental evidence consistent with orbital magnetization and spontaneous rotational symmetry breaking in trigonal Tellurium, an elemental semiconductor with a unique helical crystal structure that serves as a natural platform for investigating orbital effects. Detailed angular dependent linear and nonlinear magnetotransport measurements, supported by symmetry-guided Boltzmann transport analysis, support the interpretation of coexistence of current-induced spin polarization and orbital magnetization. With the goal of disentangling the interplay between spin and orbital degrees of freedom through electrostatic gating, this work establishes a general framework for understanding orbital magnetization in chiral crystals and beyond, paving the way for its utilization in orbitronics and spintronics.

## Introduction

The orbital degree of freedom, a fundamental component of electron angular momentum, has long been overshadowed by its spin counterpart. However, recent theoretical and experimental advancements have reignited interest in orbital effects, revealing their profound influence on material properties. Phenomena such as the orbital Hall effect[1–3], intrinsic planar Hall effect[4], chiral orbital current[5], and orbital Chern states[6] have demonstrated the rich physics arising from orbital dynamics. Moreover, Berry phase physics has established that orbital angular momentum and magnetization are intrinsic properties of electronic wavefunctions in solids[7], providing a unified framework for understanding these effects.

Despite these advancements, clear observation of orbital effects has been hindered by the inherent entanglement between the orbital and spin degrees of freedom in materials with spin-orbit coupling (SOC)[2], thus largely limited to light-element materials. Moreover, it has been shown that in several previously studied systems where the observations were attributed solely to spins, there are in fact substantial contributions from orbitals[8–10]. While orbital angular momentum can directly impact material properties through interaction with magnetic fields via Zeeman-like coupling even without SOC[4], the precise mechanisms by which orbital magnetization contribute to magnetotransport in the presence of SOC should be carefully elucidated.

Symmetry and its breaking play a crucial role in condensed matter physics. Various spatial and time-reversal symmetries have enabled the discovery of new topological phases through protected features in electronic band structures[11–14]. Conversely, symmetry breaking often leads to novel emergent phenomena[15–17], opening new frontiers in quantum materials research. One notable example is structural chirality, a distinct form of spatial inversion symmetry breaking, which has emerged as a focal point of studies[18–20]. In chiral materials, two distinct charge-spin conversion effects have been widely studied: the collinear Rashba-Edelstein effect in conducting chiral crystals[21–28] and chirality-induced spin selectivity in insulating chiral molecules[29–31]. In both cases, there are tantalizing clues that orbital polarization plays essential roles[32–38]. However, despite significant progress in understanding spin polarization in chiral materials, direct and systematic experimental identification of orbital magnetization arising from structural chirality remains difficult. Moreover, prior studies in chiral crystals have largely focused on current and field configurations aligned with the helical axis, where the ideal three-fold rotational symmetry of the chiral chain is typically assumed, and off-chain (transverse) magnetization components are forbidden by symmetry.

Here, we report systematic experimental evidence consistent with presence of orbital magnetization in Tellurium (Te) from a comprehensive set of angle-resolved magnetotransport measurements on pre-patterned 'L'-bar devices. The results point towards distinct features of orbital magnetization that coexist with spin polarization. Notably, we argue that our data is consistent with the presence of an unexpected orbital magnetization component perpendicular to the helical chains. Moreover, the relative contributions of the spin and orbital effects could be tuned by electrostatic gating, enabling precise control and detailed study of their interplay with varying chemical potential. Our theoretical framework elucidates a general relationship between the electronic structure of Te and its spin polarization and orbital magnetization. Within the minimal symmetry-based Boltzmann transport framework, we are able to qualitatively explain the experimental trends as an interplay and competition between the current-induced spin polarization and orbital magnetization. We argue that the finite magnetization component perpendicular to the helical axis is a natural consequence of symmetry reduction due to extrinsic or intrinsic mechanisms. Taking together, our experimental and theoretical findings help advance the fundamental understanding of orbital dynamics in chiral materials and highlight the ubiquity of orbital effects.

**Magnetotransport signatures of orbital magnetization**

Te crystals exhibit $D_3$ symmetry, comprising helical chains of covalently bonded atoms, arranged into a hexagonal close-packed structure via inter-chain van der Waals interactions. The right-handed crystal structure (space group $P3_121$) is illustrated in Fig. 1a. The 'L'-bar device structure (Fig. 1b), with two arms aligned along different crystalline axes, enables thorough examination of all possible configurations of current and magnetic field orientations relative to the chiral crystal axis within a single device. Single-crystal Te flakes are synthesized through hydrothermal growth[39], and then transferred onto $Si^{++}/SiO_2$ wafers. A flake is identified and patterned into an 'L'-bar, followed by another lithography step to define the metal electrodes. The fundamental ($\omega$) and second harmonic ($2\omega$) voltages are measured in both longitudinal and transverse configurations. The sample can be rotated *in situ* over 360° in two orthogonal directions as illustrated in Fig. 1b, allowing for measurements with the magnetic field oriented within the sample plane (red arrow) or out of the plane (green arrow). This setup enables the determination of eight components of longitudinal and transverse resistances: $R_{zz}^{\omega}$, $R_{xx}^{\omega}$, $R_{xz}^{\omega}$, $R_{zx}^{\omega}$, $R_{zz}^{2\omega}$, $R_{xx}^{2\omega}$, $R_{xz}^{2\omega}$, and $R_{zx}^{2\omega}$. In the following analysis, we focus on the four longitudinal components, and all data shown are taken at temperature of 2 K. In addition, using the degenerately doped Si as a back gate, the above measurements can be performed at various Fermi level at the top of the valence band, see Fig. 1c. As will be demonstrated below, this tunability allows us to control the extent of competition between the spin and orbital components. Current-induced spin polarization along the helical axis in Te due to the collinear REE is illustrated in Fig. 1d. For the orbital contribution, our model predicts orbital magnetization with two orthogonal components, $\boldsymbol{M}_{orb,z}$ along and $\boldsymbol{M}_{orb,x}$ perpendicular to the helical axis (Fig. 1e). The stretched helix crystal structure of Te (with lattice constant $c > a$) can exhibit a significant orbital moment in the transverse direction when the system's symmetry is lowered from its innate $D_3$ crystal symmetry that would otherwise prohibit it. The linear-response resistivities along *c*- and *a*-axis exhibit similar magnitudes and temperature dependence, regardless of whether the Fermi level lies within the bandgap or inside the valence band. Hall measurements also reveal nearly identical carrier densities along these two axes (Extended Data Fig. 1 and reference therein[40–42]). These common properties ensure valid direct comparison of linear and nonlinear magnetotransport stemming from spin polarization and orbital magnetization along the two axes, which rely on values of $R_{zz}^{\omega(2\omega)}$ and $R_{xx}^{\omega(2\omega)}$. The linear-response magnetoresistance (MR) was calculated from the symmetric component of fundamental (first harmonic) voltage response to an a.c. bias current with varying magnetic field. To quantify the nonlinear contribution, the quantity $\delta^{NL} = \frac{2R^{2\omega}}{R^{\omega}}$ is employed; the antisymmetric component of which with respect to the magnetic field excludes finite $R^{2\omega}$ under zero magnetic field (Extended Data Fig. 1e-f) and any contributions from trivial MR, nonlinear resistivity, and ordinary Hall effect due to misalignment (see Methods).

Figure 2a shows the angular dependence of the linear MR along the helical axis ($R_{zz}^{\omega}$) under an in-plane magnetic field ranging from 2 T to 12 T in increment of 2 T. The data was obtained from device D1 at the natural doping level of the as-grown sample. The zero-field $R_{zz}^{\omega}$ is represented by the horizontal green dotted line for reference. The linear MR exhibits 180° periodicity with respect to the magnetic field, with negative MR observed when the field aligns with the helical axis. The resistance minima, marked by the vertical gray dashed lines, reveals an experimental misalignment of approximately 5° with respect to the helical axis. However, this misalignment is systematic and independent of the magnetic field and gate voltage, thus has no impact on the subsequent discussions. We note that there is no discernible difference between the resistances with the magnetic field parallel ($\phi \approx 360°$) or antiparallel ($\phi \approx 180°$) to the current direction.

The angular dependence of $\delta_{zz}^{NL}$ measured in D1 is shown in Fig. 2b, which exhibits a 360°-periodicity with a magnitude that increases monotonically but nonlinearly with the magnetic field. Notably, at low magnetic fields, the angular positions of the maxima (minima) deviate significantly from 360° (180°) expected from current-induced collinear spin polarization. As the field strength increases, the maximum (minimum) approaches 360° (180°) and the angular shift from the 360° (180°) orientation diminishes toward zero. The angular shift at low fields is more apparent, as shown in Fig. 2g, which presents the quantity normalized by the magnetic field, $\frac{\delta_{zz}^{NL}}{\mu_0 H}$; an angular shift of approximately 45° at 2 T is evident. For the *a*-axis, Fig. 2c presents the angular dependence of $\delta_{xx}^{NL}$, along with its normalized data shown in Fig. 2h. The nonzero $\delta_{xx}^{NL}$ may arise from anisotropic band dispersion such as trigonal warping[43] in the valence band. Interestingly, at 2 T, $\delta_{xx}^{NL}$ exhibits wide plateaus centered at 360° and 180°, with a normalized magnitude comparable to $\delta_{zz}^{NL}$. As the magnetic field strength increases, additional side peaks emerge, and the normalized magnitude decreases significantly, making $\delta_{xx}^{NL}$ much smaller than $\delta_{zz}^{NL}$ at high fields. Although pronounced features consistent with collinear current-induced spin polarization are observed in D1, the angular shift in $\delta_{zz}^{NL}$ at small fields and complex magnetic field dependence of $\delta_{xx}^{NL}$ cannot be explained by spin polarization alone. The comparable magnitudes of $\delta_{zz}^{NL}$ and $\delta_{xx}^{NL}$ at 2T strongly suggest an additional underlying contribution to $\delta^{NL}$.

In our experimental configuration, the $\phi$-angle dependence of $\delta^{NL}$ tracks the direction of the total current-induced magnetization $\boldsymbol{M}$ (spin + orbital) as the sample is rotated in plane by over 360°. In the presence of collinear REE, the spin polarization is locked to the momentum direction, and the intrinsic *C*~3~ rotational symmetry of trigonal Te crystal forbids any net transverse magnetization[33,36,44]. Therefore, to account for the observed angular shift of $\boldsymbol{M}$, a reduction of rotational symmetry is required. Such symmetry reduction can naturally arise through either extrinsic or intrinsic mechanisms without explicit structural symmetry breaking. Extrinsically, because the helical chains lie within the plane of the flake interfaced with the substrate, interfacial strain or gate-induced electric fields can break rotational symmetry at the interface. Intrinsically, beyond the single-particle picture, electron–electron interactions may lead to spontaneous symmetry breaking and the formation of charge-ordered phases, as suggested by recent scanning probe measurements[45]. Once the rotational symmetry is reduced, the orbital magnetization is allowed to acquire a transverse component, while the spin polarization remains locked to the momentum direction due to broken mirror symmetry. Importantly, owing to symmetry constraints, the angular shift of the total magnetization provides an unbiased means to disentangle spin and orbital contributions without requiring knowledge of the absolute sign of either component. Consistent with this picture, our minimal symmetry-guided model incorporating reduced rotational symmetry qualitatively reproduces the experimental observations. In addition, an angular shift of up to 15 degrees in nonlinear conductivity along helical axis has been widely seen in thin Te flakes[26,36,44]. Although this shift is rarely discussed and is often attributed to imperfect sample alignment because a net transverse magnetization is forbidden by symmetry, our analysis provides a consistent, symmetry-allowed framework that accounts for this frequently observed behavior.

Figures 2d-f present the angular dependence of $R_{zz}^{\omega}$, $\delta_{zz}^{NL}$, and $\delta_{xx}^{NL}$ measured on device D2 without an applied gate voltage under in-plane rotation of the magnetic field from 1 T to 16 T, in increment of 1 T. These results exhibit contrasting behavior compared to those of D1. Notably, the angular position of the $R_{zz}^{\omega}$ minimum shifted from $\phi \approx 180^{\circ}$ (or $\phi \approx 360^{\circ}$) and increased approximately linearly with the magnetic field, as indicated by the gray dashed lines. Note that this field-enhanced angular shift is absent in Fig. 2a. Similarly, field normalized $\delta_{zz}^{NL}$ and $\delta_{xx}^{NL}$ in Figs. 2i and 2j, respectively, also display linear field dependence, with their angular shifts increasing as the field strength rises (indicated by gray dashed lines). However, the magnitude and shifted angle of $\delta_{zz}^{NL}$ are significantly greater than those of $\delta_{xx}^{NL}$, consistent with the smaller angular shift in $R_{xx}^{\omega}$ (Extended Data Fig. 2). Specifically, the onset of the $\delta_{zz}^{NL}$ angular shift is approximately 20° left of 180° (or 360°), whereas $\delta_{xx}^{NL}$ exhibits only a slight deviation from these angles. The observations from D2 clearly extend beyond the framework of collinear spin polarization, suggesting the presence of a polarization/magnetization component perpendicular to the helical axis that competes with the collinear spin polarization. This component couples to the external magnetic field linearly, becoming more pronounced as the field increases and leads to the enhanced angular shift away from the helical chain direction. The a.c. current dependences of $R^{\omega}$ and $\delta^{NL}$ of D1 and D2 are shown in Extended Data Figs. 3 and 4, respectively, suggesting the presence of additional higher-order nonlinear contributions (see Methods).

The contrasting behaviors between D1 and D2 can be attributed to the different effective SOC strengths due to the different native doping levels in the two samples, which was corroborated by quantum transport measurements. Quantum interference effects related to SOC in Te are known to exhibit significant tunability via electrostatic gating[46], and in our devices we observed a crossover from weak localization (WL) to weak anti-localization (WAL) upon electrostatic gating (Extended Data Fig. 5). Resistivity versus temperature and perpendicular field MR measurements on D1 and D2 are shown in Extended Data Fig. 6. In D1, the out-of-plane field MR shows a flat response near zero field, indicating an intermediate effective SOC strength consistent with MR in the transition regime between WL and WAL. In contrast, D2 exhibits a negative MR at zero gate voltage, consistent with WL and weak spin-orbit scattering. The larger effective spin-orbit scattering in D1 is expected to induce collinear spin polarization, as evidenced by Fig. 2b at high fields, whereas in D2 the orbital effects can directly manifest in magnetotransport because of the weak spin-orbit scattering. Based on our experimental observations, along with the symmetry considerations and the WL/WAL analysis, we argue that the observed angular shift of $\delta_{zz}^{NL}$ and $\delta_{xx}^{NL}$ in D2 is evidence for a pronounced orbital contribution in current-induced magnetization, which contains a finite off-chain (transverse) component. We emphasize that although a magnetic field along the helical axis may induce energy dispersion shifts[47], they would not lead to a shift in angular dependence in either linear or nonlinear MR (Fig. 4a).

**Gate tunability**

To further demonstrate the veracity of the picture described above, we employed electrostatic gating to modulate the electronic properties of D2 and examined the evolution of WL behavior with the Fermi level in the same sample. The tuning covered a wide range of chemical potential below the valence band maximum. Figure 3a presents the normalized $R_{zz}^{\omega}$ under different gate voltages where a higher negative voltage corresponds to a chemical potential residing deeper inside the valence band. Notably, the angular shift of the resistance minima vanishes at sufficiently large negative gate voltages, closely aligning with the transition from WL to WAL, i.e. strengthening of the spin polarization contribution as the magnitude of the negative gate voltage increases (Extended Data Fig. 6). Mapping of $\delta_{zz}^{NL}$ with the gate voltage (Fig. 3b) illustrates a clear transition between the orbital effects dominated and spin polarization dominated regions, marked by the vanishing angular shift from the 180º (antiparallel) and 360º (parallel) field orientation (indicated by the white dashed lines). The gate voltage mapping of $\delta_{xx}^{NL}$ is shown in Extended Data Fig. 7, which demonstrates the same transition. The enhancement of SOC deeper in the valence band induces stronger collinear spin polarization, while the pure orbital effect is quenched due to enhanced SOC as the chemical potential moves away from the band edge, resulting in the maxima (minima) of $\delta_{xx}^{NL}$ appearing at $\phi \approx 360^{\circ}$ ($\phi \approx 180^{\circ}$). It is interesting to note that, near the band edge, the magnitude of $\delta_{zz}^{NL}$ exceeds that inside the valence band (due to spin polarization) and continues to increase as the gate voltage increases (Extended Data Fig. 8). This suggests that the orbital effects lead to greater nonlinear MR than spin alone, further highlighting that additional spin polarization may have originated from the conversion of orbital angular momentum in the presence of SOC. These findings highlight the high tunability of the interplay between orbital and spin effects, broadening potential applications in orbitronics and spintronics.

To conclude the experimental characterization of the samples, we also examined the case of an out-of-plane magnetic field, where nonlinear MR has been observed in both bulk and 2D Te[27,48]. Figures 3c and 3d present the angel $\theta$ dependence of $\delta_{zz}^{NL}$ measured on D2 under out of sample plane tilt at different gate voltages and applied field. The $\delta^{NL}$ in the presence of an out-of-plane magnetic field is consistently captured, with a magnitude comparable to that seen in in-plane sample rotation. A hump at $\theta = 0^{o}$ ($\phi = 270^{o}$) and a dip at $\theta = 180^{o}$ ($\phi = 90^{o}$) are observed, consistent with nonzero $\delta^{NL}$ at those positions due to angular shift under in-plane rotation at $V_{gate} = 0$ V. Notably, as the angular shift vanishes when the Fermi level is tuned to spin polarization dominant region (deeper in the valence band), the curve reverts to a simple sin($\theta$)-like behavior with respect to out-of-plane magnetic field, aligning with previous observations in Te[27,48]. Taken together, our gate tuning measurements significantly extend the understanding of the magnetotransport properties in this material

**Discussions**

To help interpret the experimental results, we perform a semiclassical Boltzmann transport calculation to extract both the linear and nonlinear conductivities. Full details of the calculation are provided in the Supplementary Information and References therein[36,38,49–69]. Within linear response to magnetic field, our model takes the general form:

$$\varepsilon(\boldsymbol{k}) = \varepsilon_0(\boldsymbol{k}) - \mu_B B_{\parallel} \cos(\phi) \left(s_z(\boldsymbol{k}) + m_z(\boldsymbol{k})\right) - \mu_B B_{\parallel} \sin(\phi) \left(s_x(\boldsymbol{k}) + m_x(\boldsymbol{k})\right)$$

Here, $\varepsilon_0(\boldsymbol{k})$ denotes the band dispersion without a magnetic field, $\mu_B$ is the Bohr magneton, and the latter terms represent the Zeeman coupling to an in-plane magnetic field of magnitude $B$ oriented along (sin($\phi$), 0, cos($\phi$)). Total magnetization ($\boldsymbol{M}$) thus includes the summation of both orbital ($\boldsymbol{m}(\boldsymbol{k})$) and spin ($\boldsymbol{s}(\boldsymbol{k})$) contributions which we define below.

In Figs. 4a-f, we present results for three different model scenarios that help us substantiate proposed experimental interpretation of competing spin and orbital polarizations. First, in Figs. 4a, and 4d, we demonstrate that spin texture alone, i.e. the collinear spin model, is insufficient to account for the experimentally observed trends – the resistance curves' maxima/minima are rigidly locked along the collinear and anti-collinear directions. This observation is expected due to the original rotational symmetry of the model. Second, we conduct a symmetry analysis of orbital magnetization and find that only an orbital magnetization component that is odd in $k_z$ can explain the experimentally observed shift in the minima of the resistance (Figs. 4b and 4e). Here, we use a parameter $\alpha$ as a tuning parameter, such that $\alpha = 0$ corresponds to a situation in which orbital magnetization is absent and $\alpha = 1$ corresponds to purely orbital magnetization with no spin magnetization. This conclusion is further supported by a toy model of a helical chain with three distinct atoms per unit cell that break rotational symmetry of the chain, which validates the symmetry-based argument (Figs. 4c and 4f). In the Supplementary Information we review existing models of the band structure of Te and find that they either yield no orbital magnetization or the magnetization lacks the $k_z$ dependence on the crystal momentum and end up with the collinear result.

Separately, given that the surface of Te flake breaks inversion symmetry, a surface-induced Rashba SOC can generate a tangential spin texture with the same symmetry as a transverse orbital magnetization, and could therefore, in principle, produce an angular shift (see discussion in Supplementary Information). However, such a scenario is inconsistent with our observations. Firstly, a surface Rashba SOC effect should decrease with increasing the thickness of the sample (a naive scaling analysis would predict the strength of Rashba SOC decaying inversely with the thickness of the sample), whereas in our case, normalized nonlinear MR does not show strong thickness dependent (given the thickness of D1 is 80 nm and for D2 is 44 nm). Moreover, previous work[70,71] has found an increase in Rashba SOC at doping away from the Te valence-band edge. However, our experimental observation shows that deep in the valence band, the bulk radial SOC surpass the surface effect. As a result, if at the band edge the surface SOC is already larger than the bulk radial SOC, it will always be a leading order effect independent of the doping and the angular shift will show opposite trend with doping.

Taken together with experimental results, the theoretical calculations of Fig. 4 consistently support the interpretation of coexisting current-induced spin polarization and orbital magnetization, with orbital contribution being most pronounced when SOC is weak. In chiral materials such as Te, orbital magnetization differs from spin polarization by having two orthogonal components, $\boldsymbol{M}_{orb,z}$ along helical axis and $\boldsymbol{M}_{orb,x}$ perpendicular to it once the three-fold rotational symmetry is broken, as illustrated in Fig. 1e. When the SOC is strong and spin effects dominate, current-induced spin polarization $\Delta\boldsymbol{S}$ (Fig. 1d) overwhelms $\boldsymbol{M}_{orb}$. In this regime, the magnetic field both (1) couples to $\Delta\boldsymbol{S}$ to induce $\delta^{NL}$, and (2) amplifying $\Delta\boldsymbol{S}$ by creating an imbalance in spin occupation states. The angular shift at low magnetic fields (Fig. 2b) reflects the mixing of $\Delta\boldsymbol{S}$ and $\boldsymbol{M}_{orb,x}$; as the magnetic field increases, $\boldsymbol{M}_{orb,x}$ becomes negligible comparing to $\Delta\boldsymbol{S}$, resulting in collinear maxima and minima. In orbital magnetization dominate regime, the ratio between its components along and perpendicular to the helical axis evolves with increasing magnetic field (Fig. 2e) while remaining constant for different current bias (Extended Data Fig. 4b). We emphasize, however that based on symmetry analysis, we cannot uniquely determine that the origins of the mechanism are due to an orbital effect. Specifically, if a spin polarization along non-collinear direction was present (with appropriate momentum symmetry), then similar behavior could be theoretically reproduced as well. In that case, however, it would point towards a non-radial spin structure in Te in significant disagreement, both for bulk and thin-film systems, with prior experimental and theoretical evidence[21–24,26,57,72].

## Conclusion

A comprehensive set of linear- and nonlinear response magnetotransport measurements covering all relative orientations of applied magnetic field and bias current with respect to the helical crystalline axis was conducted on trigonal Te, a chiral material with a distinct helical crystal structure. The results revealed experimental evidence naturally interpreted as signatures of orbital magnetization, specifically, the coexistence of current-induced spin polarization and two orthogonal components of orbital magnetization. By disentangling their contributions to magnetotransport via electrostatic gating, we identify distinct linear- and nonlinear-response behaviors that align with theoretical predictions. These findings provide deeper insight into the interplay between spin and orbital degrees of freedom in chiral materials. Our results establish Te as an excellent platform for studying the tunable orbital magnetization with rich spin structures.

**Acknowledgements**

We acknowledge fruitful discussions with Oskar Vafek, Dimitry Chichinadze, Jaychandran Padayasi, and Zhengguang Lu. We thank Leonid Golub, Dima Pesin, and Masato Sakano for valuable discussions over email. P.X. is supported by NSF grant no. DMR-1905843 and DMR-2325147. S.J. acknowledges support from Florida State University through the Quantum Postdoctoral Fellowship and the National High Magnetic Field Laboratory. C.L. was supported by start-up funds from Florida State University and the National High Magnetic Field Laboratory. The National High Magnetic Field Laboratory is supported by the National Science Foundation through NSF/DMR-2128556 and the State of Florida.

**Author contributions**

P.X. supervised the project. Z.H. and D.G. performed the magnetotransport measurement. C.N. and P.T. grew Te single crystals. Z.H., G.S., H.L., and J.G. fabricated the devices. S.J. and C.L. performed the calculations. All authors discussed the results and wrote the paper.

**Competing interests**

The authors declare no competing interests.

**Data availability**

The data that support this study are reported in the manuscript and Supplementary Information. Source data and other data are available from the corresponding authors on request.

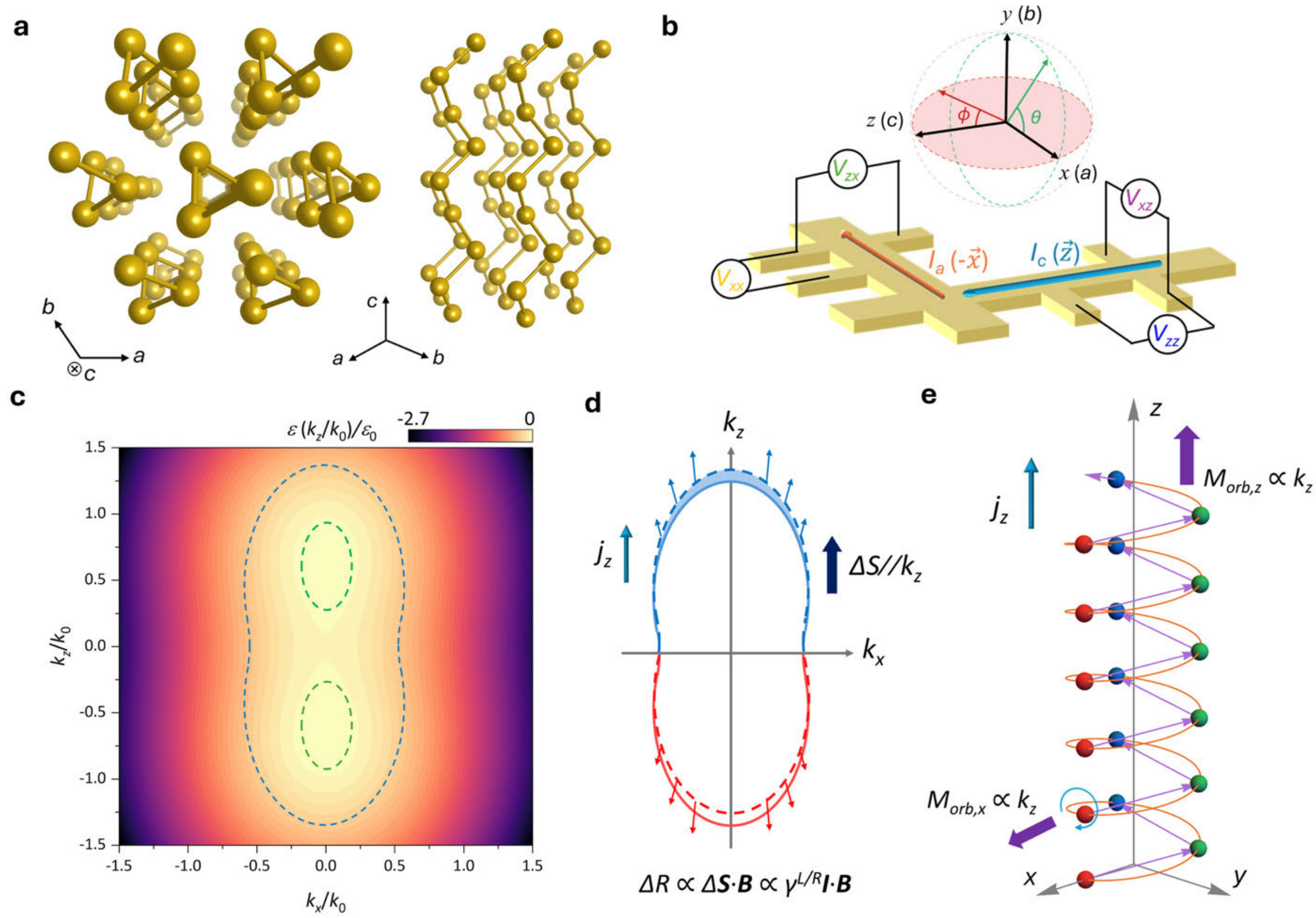

Fig. 1 | **Chiral crystal structure and experimental geometry of trigonal Te. a**, crystal structure of right-handed Te. **b**, schematic of the 'L'-bar device structure, with the two long bars aligned along the crystalline axes labeled as *c* (helical axis) and *a*. The sample can be rotated in both in-plane ($\theta$) and out-of-plane ($\phi$) directions, enabling full control of the magnetic field orientation within (red arrow) and out of (green arrow) the sample plane, where the sample plane is denoted by light red circle. **c**, calculated energy dispersion of the valence band of Te based on our theoretical model. **d**, illustration of radial spin texture and current-induced spin polarization ($\Delta\boldsymbol{S}$) in Te. The change in resistance ($\Delta R$) switches sign with variations in handedness ($\gamma^L = -\gamma^R$), current direction, or magnetic field orientation. **e**, illustration of current-induced orbital magnetization, where both components along ($\boldsymbol{M}_{orb,z}$) and perpendicular ($\boldsymbol{M}_{orb,x}$) to the helical axis, which are proportional to $k_z$.

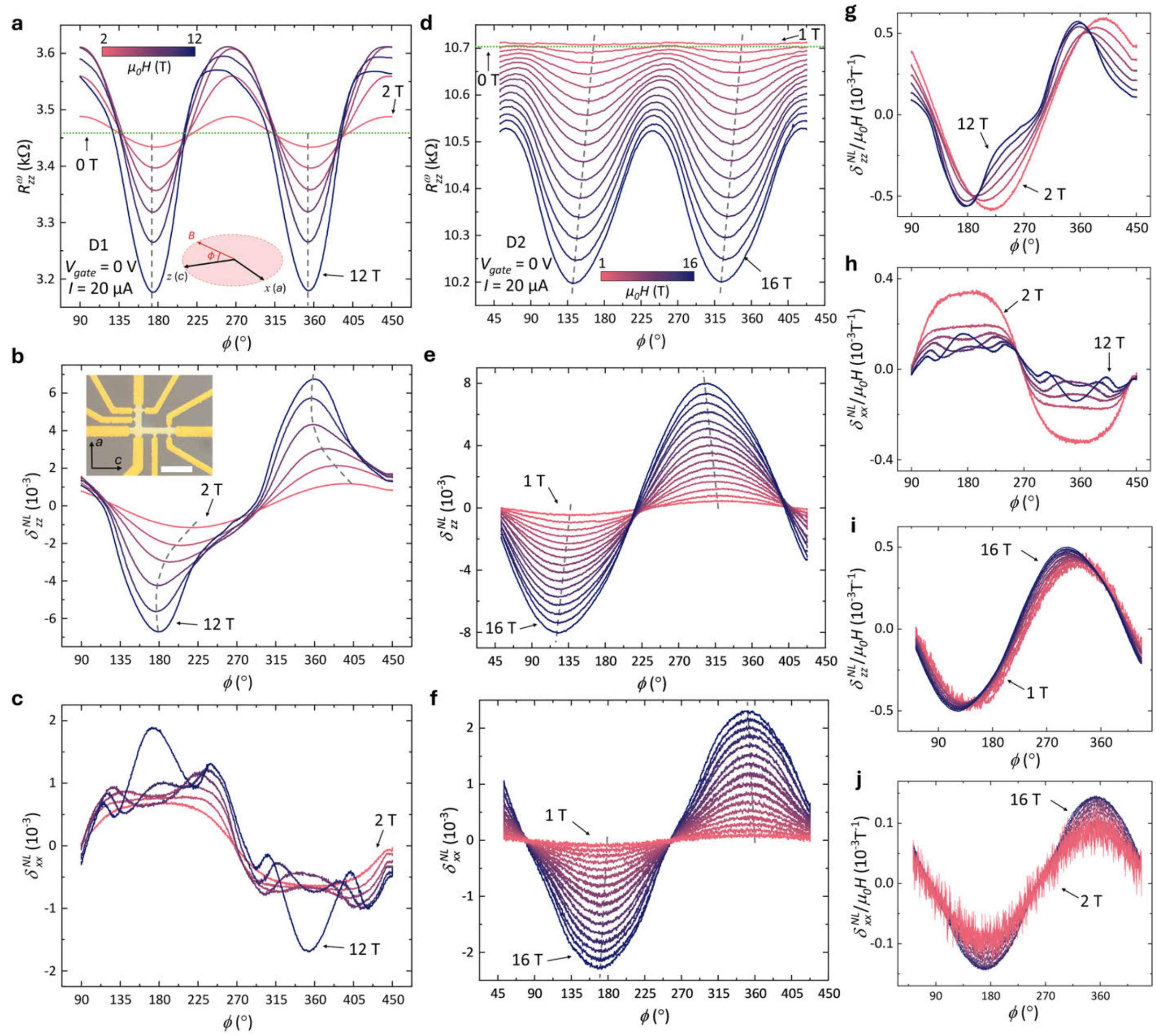

Fig. 2 | **Linear and nonlinear magnetotransport of Te. a-c**, angular dependence of linear magnetoresistance along $c$ (helical) axis, $R^{\omega}_{zz}$ (**a**), nonlinear magnetoresistance along $c$ axis, $\delta^{NL}_{zz}$ (**b**), and nonlinear magnetoresistance along $a$ axis, $\delta^{NL}_{xx}$ (**c**) under in-plane rotation of the applied magnetic field ranging from 2 T to 12 T in increments of 2 T, measured in D1 without an applied gate voltage. The current bias is 20 μA. The inset in (**b**) is an optical image of the device (scale bar: 30 μm). **d-f** angular dependence of $R^{\omega}_{zz}$ (**d**), $\delta^{NL}_{zz}$ (**e**), and $\delta^{NL}_{xx}$ (**f**) of D2 under similar measurement conditions with the magnetic fields ranging from 1 T to 16 T in increments of 1 T. $R^{\omega}$ and $\delta^{NL}$ show contrasting behavior between D1 and D2, where the observations in D2 are beyond the framework of collinear spin polarization. **g-j** magnetic field normalized data of (**b**), (**c**), (**i**) and (**f**).

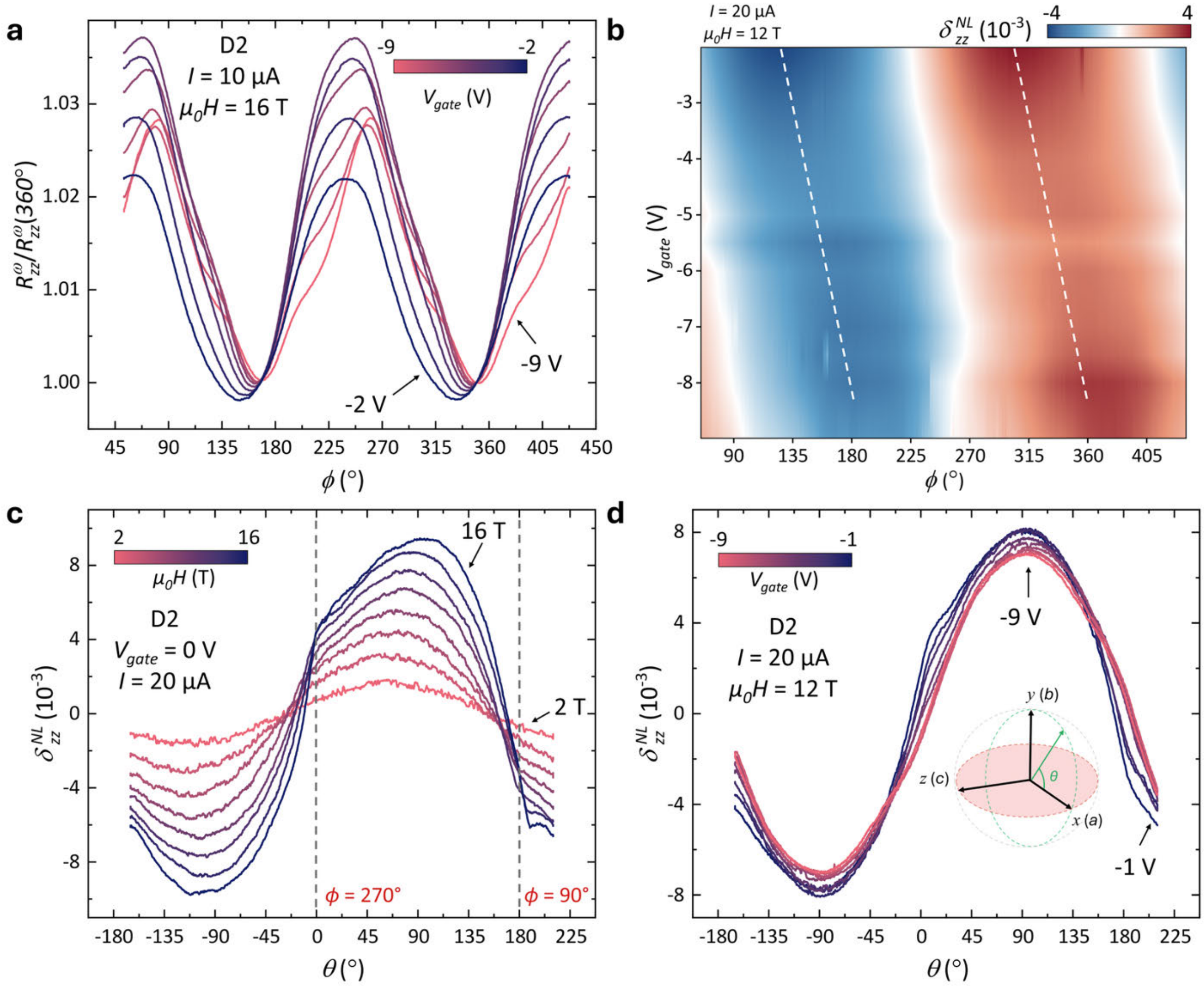

Fig. 3 | **Transition from orbital magnetization dominant to spin polarization dominant magnetotransport. a**, angular dependence of $R_{zz}^{\omega}$ for D2 under different gate voltages. The angular shift of the minimum vanishes at sufficiently low $V_{gate}$. **b**, gate voltage mapping of $\delta_{zz}^{NL}$ for $I$ = 20 μA and $\mu_0 H$ = 12 T, clearly showing the transition from the proposed orbital magnetization effect dominant region to a spin polarization dominant region, indicated by white dashed lines. **c**, $\theta$-dependence of $\delta_{zz}^{NL}$ under out-of-plane field rotation at $V_{gate}$ = 0 V, the magnetic field is incremented by 2 T. **d**, $\theta$-dependence of $\delta_{zz}^{NL}$ under different gate voltages, demonstrating the crossover between orbital and spin contributions. The hump at $\theta = 0^{o}$ and the dip at $\theta = 180^{o}$ disappear in the spin polarization-dominant region, where the angular shift also vanishes at the same applied magnetic field and current.

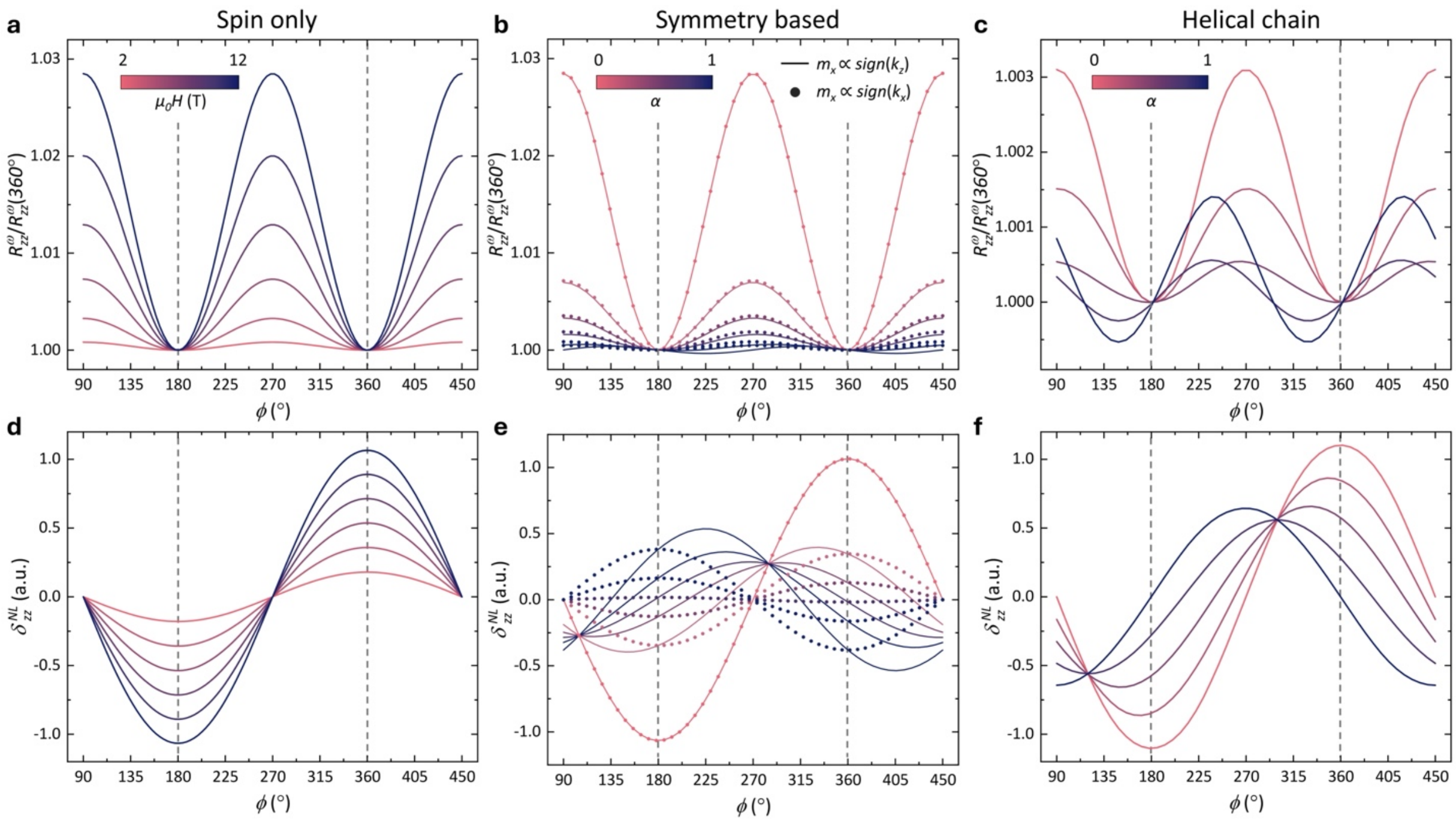

Fig. 4 | **Theoretical plots of linear and nonlinear transport:** These figures show linear and nonlinear resistivity as a function of the direction of the in-plane magnetic field, calculated using the Boltzmann semiclassical transport formalism (see Supplementary Information for the details of the calculation). **a&d** we consider an energy dispersion that includes only the Zeeman coupling to the spin texture and respects the $C_3$ rotational symmetry of the $D_3$ symmetry group of Tellurium. Here, as expected, varying the magnetic field strength does not shift the minima of the oscillations corresponding to the collinear and anti-collinear directions. **b&e**, we introduce an orbital magnetization along the *x*-direction, assumed to be odd in $\boldsymbol{k}_z$ (the figure also includes a component odd in $\boldsymbol{k}_x$, which does not reproduce the observed trend). This leads to a clear shift in the position of the minimum. **c&f.** A minimal, explicitly $C_3$ symmetry breaking, phenomenological chiral chain model for helical Te qualitatively captures the same behavior. In these four figures, $\alpha$ is a tuning parameter that controls the relative contribution of spin texture and orbital magnetization to the Zeeman energy: $\alpha = 0$ corresponds to the case with no orbital magnetization (same as the first set), while $\alpha = 1$ corresponds to a scenario with no spin polarization.

## Methods

### Sample and device fabrication

Single crystal Te were grown using a hydrothermal synthesis method. In a typical procedure, 0.09 g of $Na_2TeO_3$ (Sigma-Aldrich) and 0.50 g of PVP (Sigma-Aldrich) were fully dissolved in 33 mL of deionized water (18.2 MΩ·cm). Then, 3.33 mL of aqueous ammonia solution (28%, w/w%) (Sigma-Aldrich) and 1.67 mL of hydrazine hydrate (80%, w/w%) (Sigma-Aldrich) were added. After thorough mixing, the solution was placed into a 50 ml Teflon container, sealed in a stainless-steel autoclave, and heated at 180 °C for 30 hours in a furnace. The autoclave was then allowed to cool naturally to room temperature.

The synthesized Te flakes were transferred onto Si/$SiO_2$ wafer with degenerately doped Si and 90 nm thick $SiO_2$ using the Langmuir-Schaefer technique. The 'L'-bar device pattern was defined by photolithography, with a spin-coated layer of photoresist (AZ5214E) and etched using reactive ion etching (50 mTorr Ar, 50 W RF power, 2 mins). Electrodes were patterned in a second photolithography step, followed by the deposition of 80 nm Au using a thermal evaporator.

### Linear and nonlinear electrical measurements.

'L'-bar devices were wired to a customized sample stage made of G-10 using Pt wires and silver paint. The stage was then installed in a Physical Property Measurement System (PPMS, Quantum Design) for transport measurements at temperatures down to 2 K. The sample probe is equipped with a high-precision and low friction rotator which is driven by a DC servomotor, with the angle calculated from the motor steps and calibrated using a Hall sensor (THS118, Toshiba) for each measurement. The a.c. current was injected using Stanford Research CS580 voltage-controlled current source, synchronized with Stanford Research S860 lock-in amplifiers. The lock-in amplifiers were used to measure the fundamental and the second-harmonic voltages. Electrostatic gating is sourced by the Keithley 2450 source-measure unit.

### Linear and nonlinear magnetotransport via a.c. measurements

Linear and non-linear transport coefficients are extracted using an a.c. current. Specifically, an a.c bias current, $I = I_0 \sin\omega t$ at frequency $f = \omega/2\pi$ = 17.17 Hz, is applied to the devices (see Extended Data Fig. 1a). The resulting longitudinal voltage can be expressed as

$$\begin{aligned} V &= IR = R^{(0)}I + R^{(1)}I^2 + \cdots \\ &= R^{(0)}I_0 sin\omega t + R^{(1)}I_0^2 sin^2\omega t \\ &= R^{(0)}I_0 sin\omega t + \frac{1}{2}R^{(1)}I_0^2 + \frac{1}{2}R^{(1)}I_0^2 \sin\left(2\omega t - \frac{\pi}{2}\right) \end{aligned} \quad \text{(E1)}$$

where $R^{(n)}$ represents the $n^{th}$-order nonlinear resistance term with respect to current, higher-order terms are neglected. The fundamental ($\omega$) and second harmonic ($2\omega$) voltages components are captured by two synchronized lock-in amplifiers at phase 0° and -90°, respectively. Linear and nonlinear magnetotransport are defined as

$$R^{\omega} = \frac{V^{\omega}}{I_0} = R^{(0)},\ \delta^{NL} = \frac{2R^{2\omega}}{R^{\omega}} = \frac{2V^{2\omega}}{V^{\omega}} = \frac{R^{(1)}I_0}{R^{(0)}} \tag{E2}$$

Since linear resistance $R^{(0)}$ (e.g. intrinsic and trivial MR) is even with respect to the magnetic field, we extract its magnetic field symmetric component ($R^{\omega}$) to exclude the regular Hall effect contribution (due to misalignment of the sample orientation to the plane). In contrast, $\delta^{NL}$ arises from spin polarization and orbital magnetization, both of which couple to the magnetic field via Zeeman coupling and are odd under field reversal. Therefore, we take the magnetic field antisymmetrized $\delta^{NL}$ to isolate other effects.

**Resistivity and carrier density along different crystalline axes**

The resistivity of Te along $c$-axis ($\rho_{zz}^{\omega}$, solid line) and $a$-axis ($\rho_{xx}^{\omega}$, dashed line) measured on device D3 at different gate voltages is shown in Extended Data Fig. 1b. When no gate voltage is applied or at $V_{gate}$ = -5 V, the Fermi level lies within the bandgap. At $V_{gate}$ = -10 V, it shifts into the camelback region, and at $V_{gate}$ = -20 V, it enters the deep valence band. The resistivities exhibit similar magnitudes (within a factor of 1.5) and follow similar temperature dependencies, regardless of whether the Fermi level lies within the bandgap or inside the valence band. This observed similarity in resistivity along different crystalline axes aligns with previous results obtained from Hall bar devices[35], but contrasts with recent resistivity tensor analyses based on 'sunflower'-shaped devices, which reported significant anisotropy[36]. This discrepancy underscores the critical role of sample geometry in resistivity measurements, as variations in the electric potential distribution between 'sunflower' and Hall bar (L'-bar) configurations lead to differing results[37] and we leave a closer examination of this effect for future work.

Extended Data Figs. 1c and 1d display Hall measurements performed on device D2 at gate voltages ranging from -10 V to 0 V, along the $c$-axis and $a$-axis, respectively. The insets illustrate the sheet hole density for both axes, which exhibit similar magnitudes (in the order of $10^{12}$ $cm^{-2}$) and gate dependencies. The comparable fundamental characteristics along the helical axis and $a$-axis justify a direct comparison of $R^{\omega}$ and $\delta^{NL}$ between these two orientations.

**Nonlinear current dependence**

Based on the above discussion, $R^{\omega}$ is expected to remain constant with current, while $\delta^{NL}$ should vary linearly. However, observed current dependence measured on D1 (Extended Data Fig. 3) and D2 (Extended Data Fig. 4) otherwise, suggesting the presence of higher-order terms. These contributions can also be captured using the same measurement scheme. If the3$^{rd}$- and 4$^{th}$- order terms of $R^{(n)}$ are nonzero, Eq. E1 can be rewritten as:

$$V = IR = R^{(0)}I + R^{(1)}I^2 + R^{(2)}I^3 + R^{(3)}I^4 + \cdots$$

$$= R^{(0)}I_0 sin\omega t + R^{(1)}I_0^2 sin^2\omega t + R^{(1)}I_0^3 sin^3\omega t + R^{(1)}I_0^4 sin^4\omega t$$

$$= R^{(0)}I_0 sin\omega t + \frac{1}{2}R^{(1)}I_0^2 + \frac{1}{2}R^{(1)}I_0^2 \sin\left(2\omega t - \frac{\pi}{2}\right) + \frac{3}{4}R^{(2)}I_0^3 sin\omega t - \frac{1}{4}R^{(2)}I_0^3 sin3\omega t$$

$$+ \frac{3}{8}R^{(3)}I_0^4 + \frac{1}{2}R^{(3)}I_0^4 \sin\left(2\omega t - \frac{\pi}{2}\right) + \frac{1}{8}R^{(3)}I_0^4 \sin\left(4\omega t - \frac{\pi}{2}\right) \quad \text{(E3)}$$

It is clear from above that, $R^{(2)}$ contributes to $V^{\omega}$, and $R^{(3)}$ enters $V^{2\omega}$. Consequently, the expressions for linear and nonlinear magnetotransport become

$$R^{\omega} = \frac{V^{\omega}}{I_0} = R^{(0)} + \frac{3}{4}R^{(2)}I_0^2,\ \delta^{NL} = \frac{2R^{2\omega}}{R^{\omega}} = \frac{2V^{2\omega}}{V^{\omega}} = \frac{R^{(1)}I_0}{R^{(0)}} + \frac{R^{(3)}I_0^3}{R^{(0)}} \quad \text{(E4)}$$

**Tunable SOC probed by WL(WAL) measurement**

The spin-orbit interaction in Te exhibits high gate tunability[39], as demonstrated by WL and WAL measurements. As SOC introduces an additional phase shift, which can flip the interference from constructive to destructive, leading to a WL to WAL transition. In a typical Hall bar device, a transition from WL to WAL is observed as the Fermi level shifts from the valence band edge to deeper within the valence band (Extended Data Fig. 5). This transition indicates weak SOC near the band edge and stronger SOC in the deeper valence band.

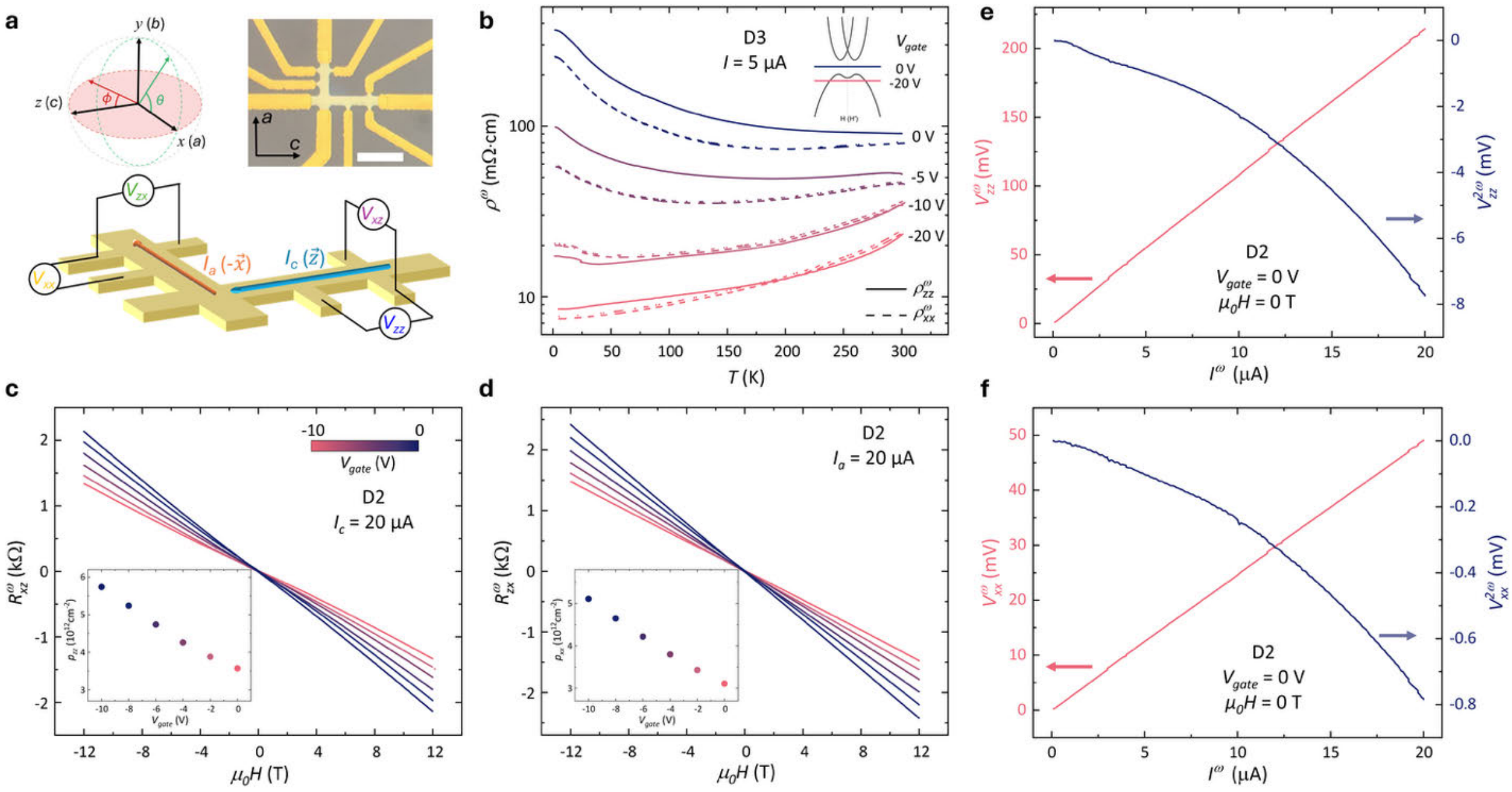

Extended Data Fig. 1 | **a**, illustration of the measurement geometry and an optical image of the device. (scale bar: 30 μm). **b**, resistivity along *c*-axis (solid lines) and *a*-axis (dashed lines) at different gate voltages measured on D3, the inset depicts a simplified band structure and indications of Fermi level. **c**, **d**, Hall resistance measured on D2 at different gate voltages along c-axis (**c**) and a-axis (**d**), with insets showing the corresponding sheet hole densities. **e**, **f**, fundamental ($V^{\omega}$) and non-zero second-harmonic ($V^{2\omega}$) voltage measured in D2 at zero field along *c*-axis (**e**) and *a*-axis (**f**).

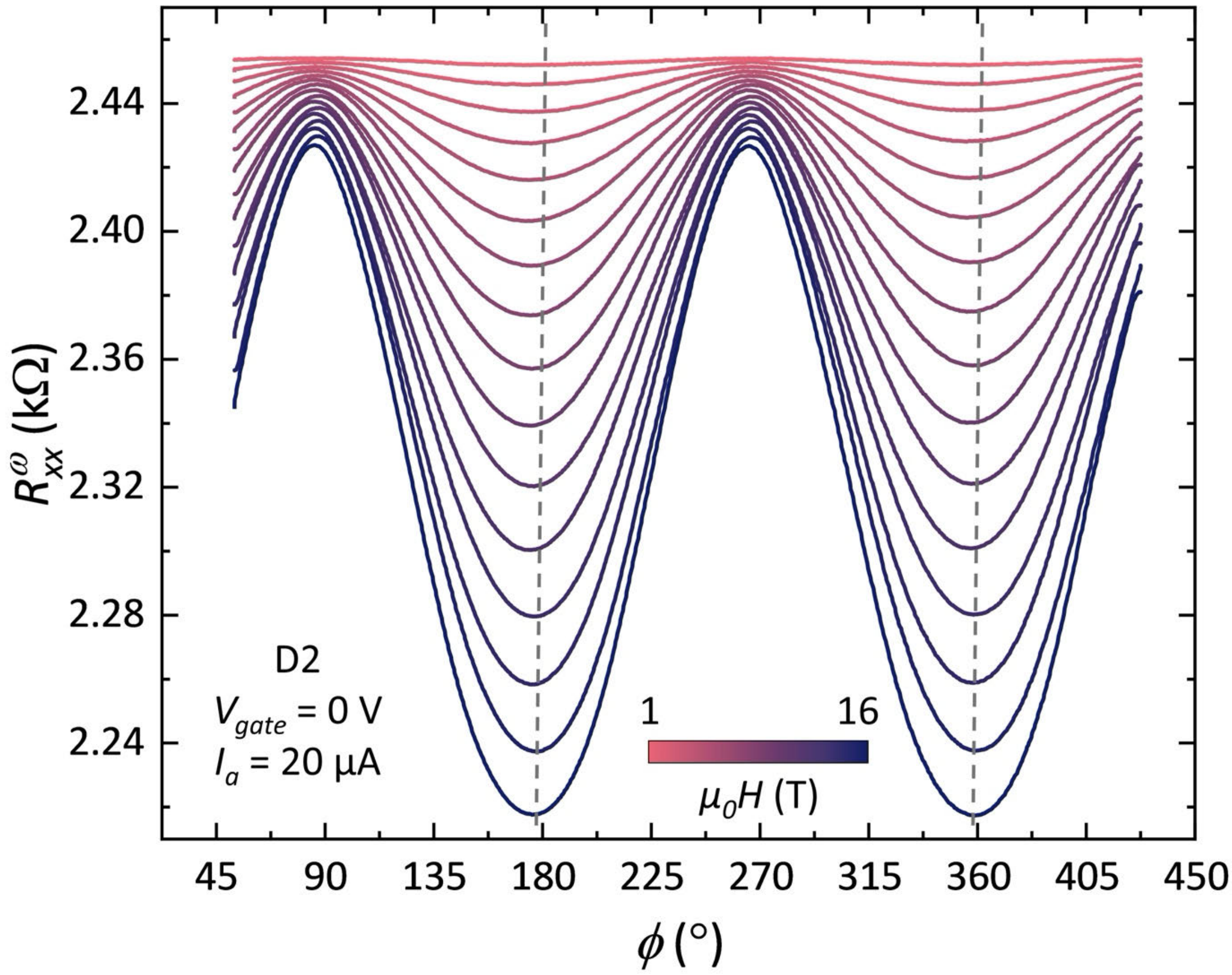

Extended Data Fig. 2 | $R_{xx}^{\omega}$ of D2 measured under in-plane rotation at different magnetic fields, compared to Fig. 2d in the main text. $R_{xx}^{\omega}$ exhibits a smaller angular shift (gray dashed lines).

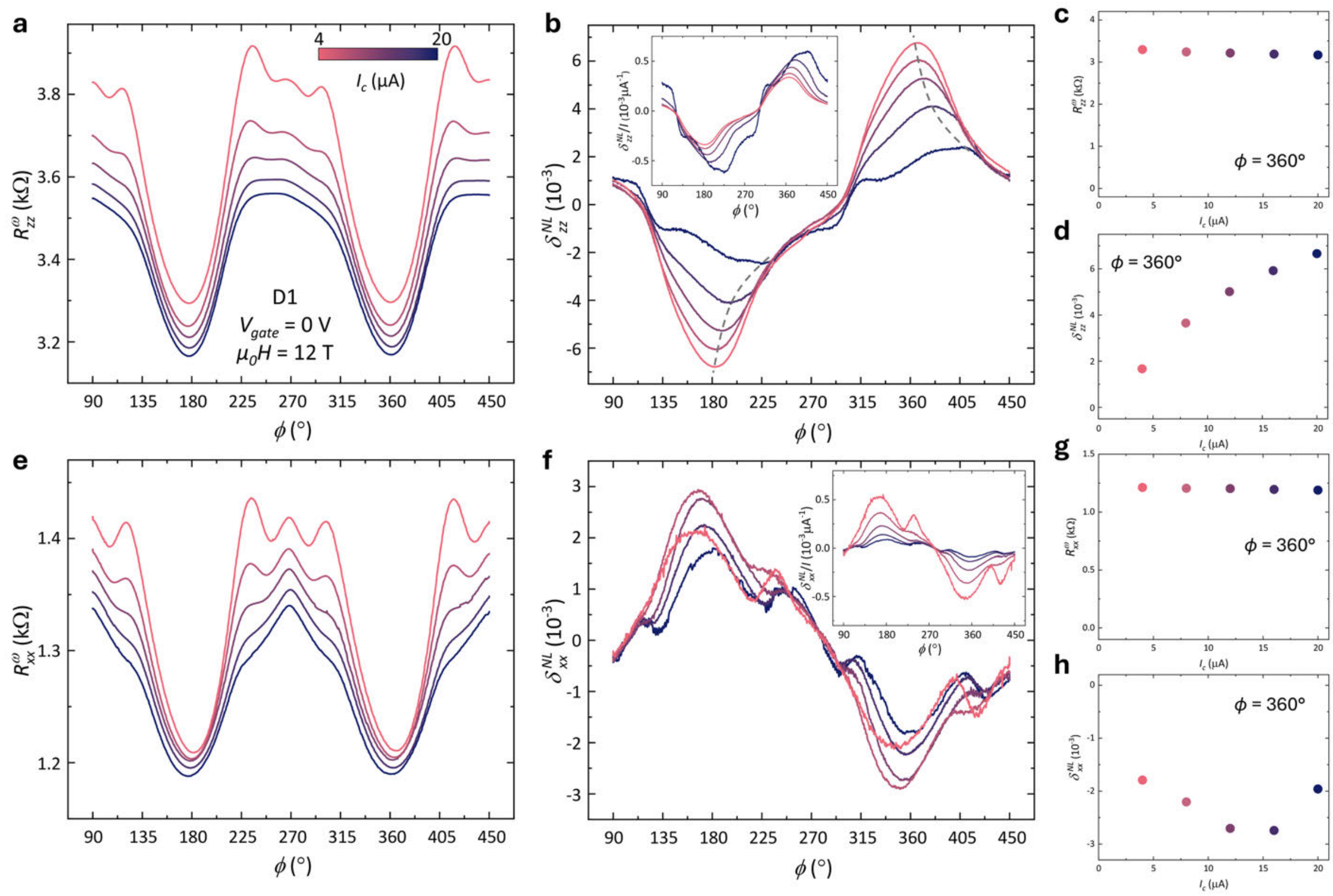

Extended Data Fig. 3 | **a**, **b**, **e**, **f**, $R^{\omega}_{zz}$ (a), $\delta^{NL}_{zz}$ (b), $R^{\omega}_{zz}$ (e), and $\delta^{NL}_{zz}$ (f) measured under in-plane rotation at different current from 4 μA to 20 μA, with an increment of 4 μA. **c, d**, **g**, **h** $R^{\omega}_{zz}$ (c), $\delta^{NL}_{zz}$ (d), $R^{\omega}_{zz}$ (g), and $\delta^{NL}_{zz}$ (h) versus current under 12 T for a fixed angle. The observed nonlinear current dependence can be attributed to high-order terms in Eq. (E4).

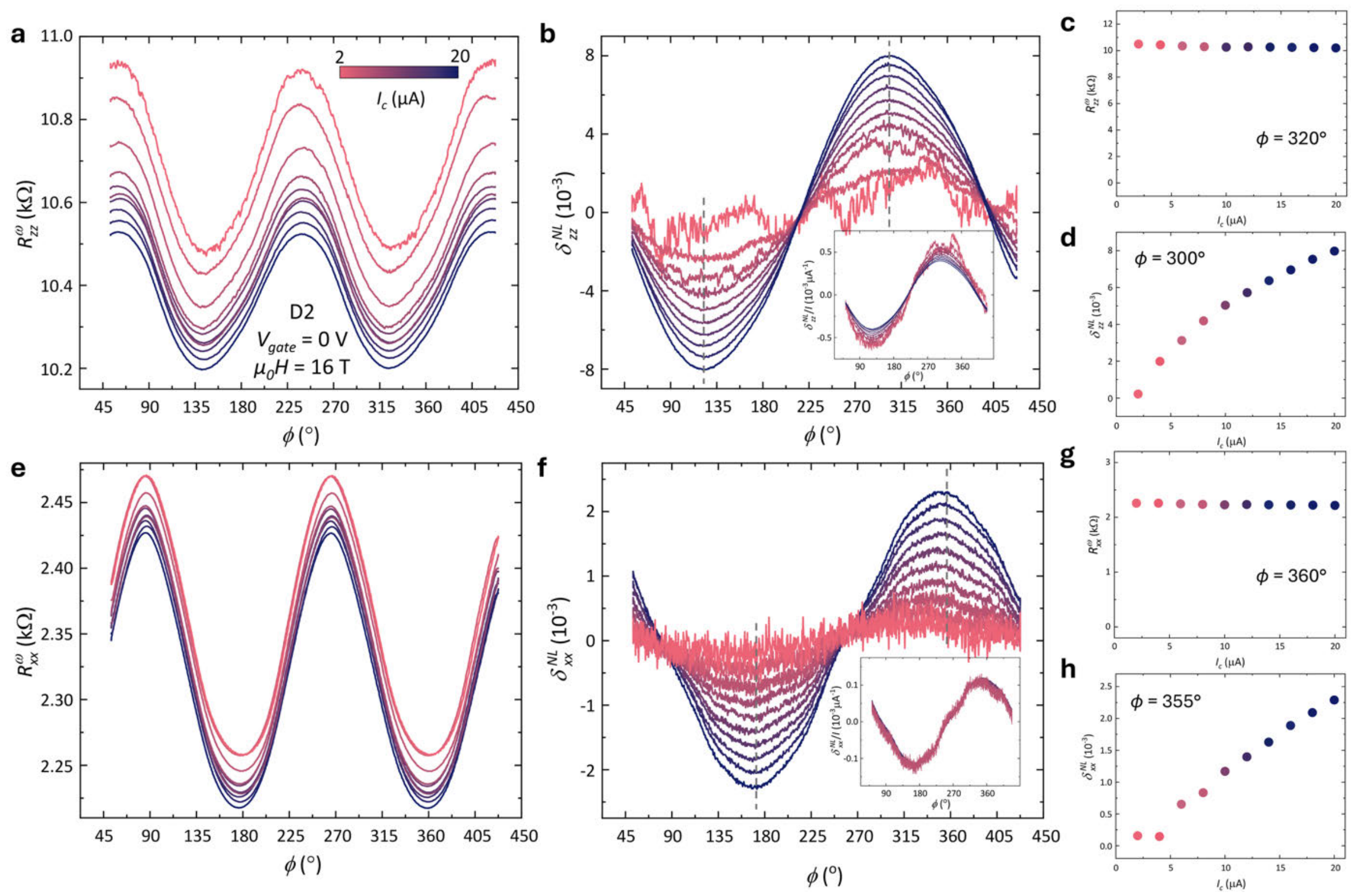

Extended Data Fig. 4 | **a**, **b**, **e**, **f**, $R^{\omega}_{zz}$ (a), $\delta^{NL}_{zz}$ (b), $R^{\omega}_{zz}$ (e), and $\delta^{NL}_{zz}$ (f) measured under in-plane rotation at different current from 2 μA to 20 μA, with an increment of 2 μA. **c, d**, **g**, **h** $R^{\omega}_{zz}$ (c), $\delta^{NL}_{zz}$ (d), $R^{\omega}_{zz}$ (g), and $\delta^{NL}_{zz}$ (h) versus current under 16 T for different fixed angles.

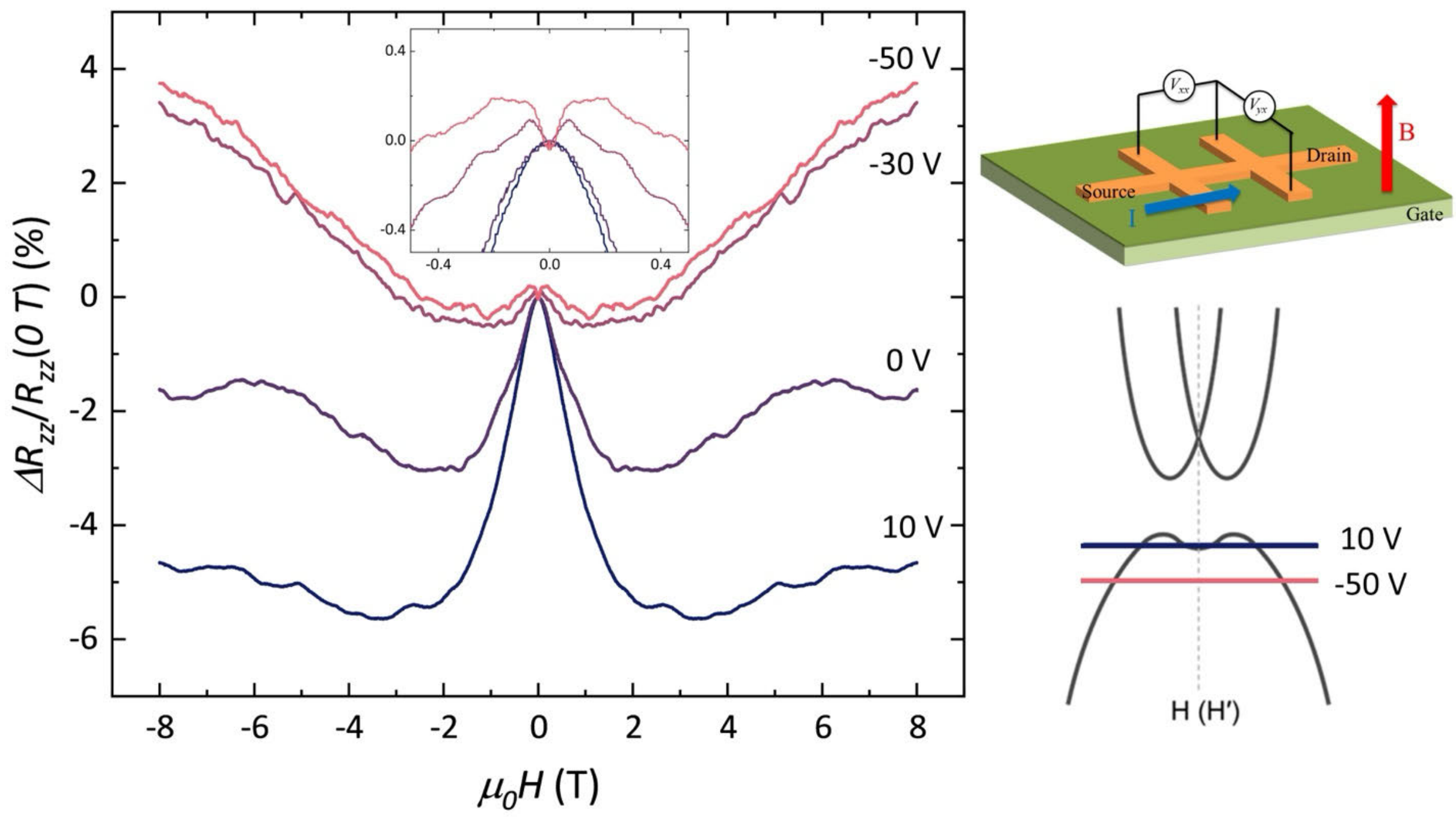

Extended Data Fig. 5 | MR measured on a typical Hall bar Te device (top right), showing a transition from WL to WAL as gate voltage decreases. The inset provides a zoomed-in view, where a clear WAL feature emerges at $V_{gate}$ = -50 V. The bottom right panel illustrates a simplified band structure diagram, depicting the Fermi level positions at different gate voltages.

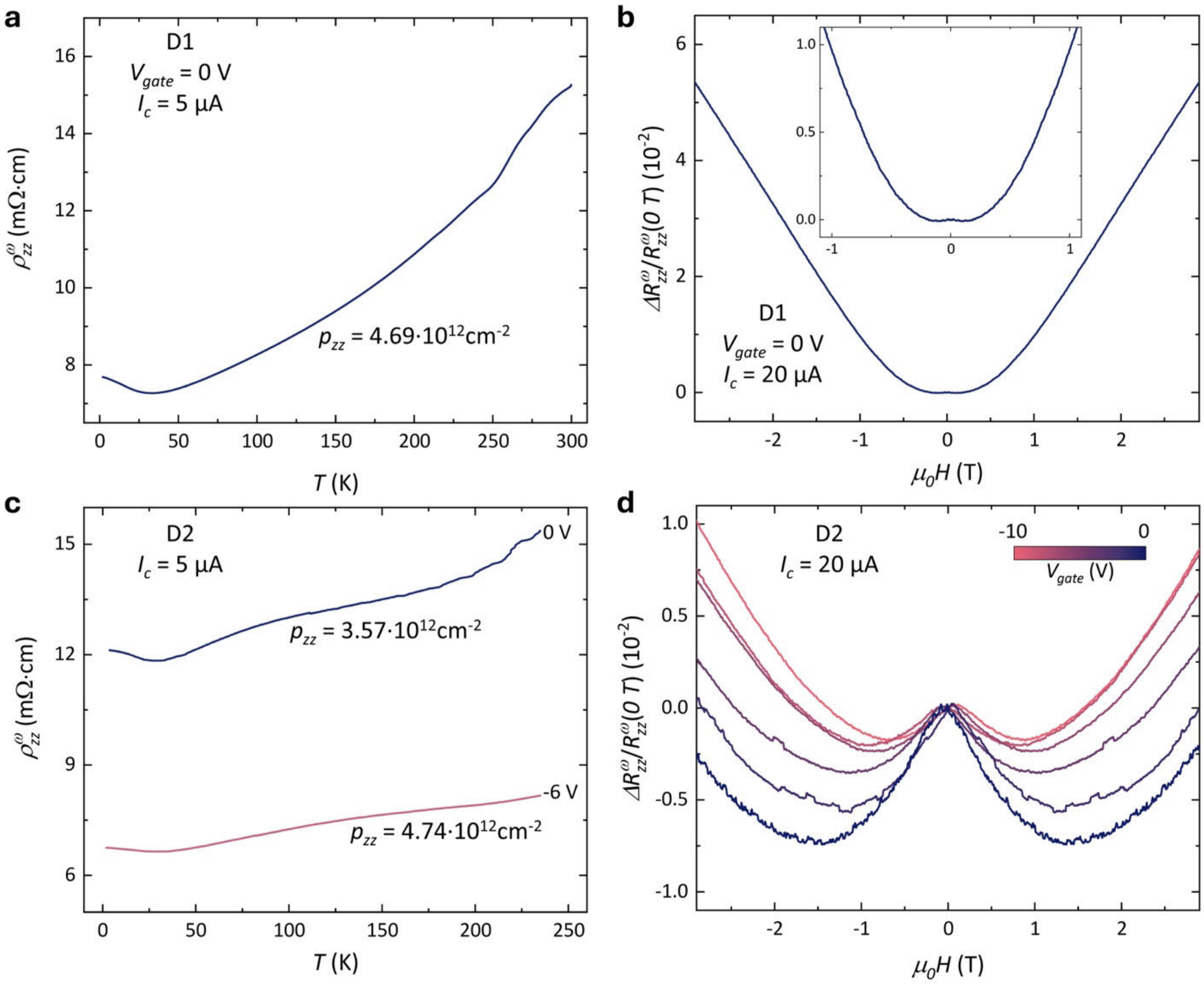

Extended Data Fig. 6 | **a**, **c** Resistivity versus temperature measurements on D1 (**a**) at $V_{gate}$ = 0 V and D2 (**b**) at $V_{gate}$ = 0 V and -6 V, confirming that the Fermi level remains within the valence band across all discussed ranges. Hall measurements indicate a similar 2D carrier density between D1 ($V_{gate}$ = 0 V) and D2 ($V_{gate}$ = -6 V). **b**, **d** Linear longitudinal MR, calculated as

$\frac{\Delta R^{\omega}_{zz}}{R^{\omega}_{zz}(0)} = \frac{R^{\omega}_{zz}(\mu_0 H) - R^{\omega}_{zz}(0)}{R^{\omega}_{zz}(0)} \times 100$**,** measured on D1(**b**) and D2 (**d**) under out-of-plane field. The inset in (**b**) provides a zoomed-in view, highlighting the MR lies in the transition region from WL to WAL, indicating strong SOC in D1. In contrast, MR measurements on D2 reveal a strong WL signature, especially at $V_{gate}$ = 0 V, indicating weak SOC. This supports that the behavior observed in Fig. 2d-f originates from orbital effects rather than spin polarization.

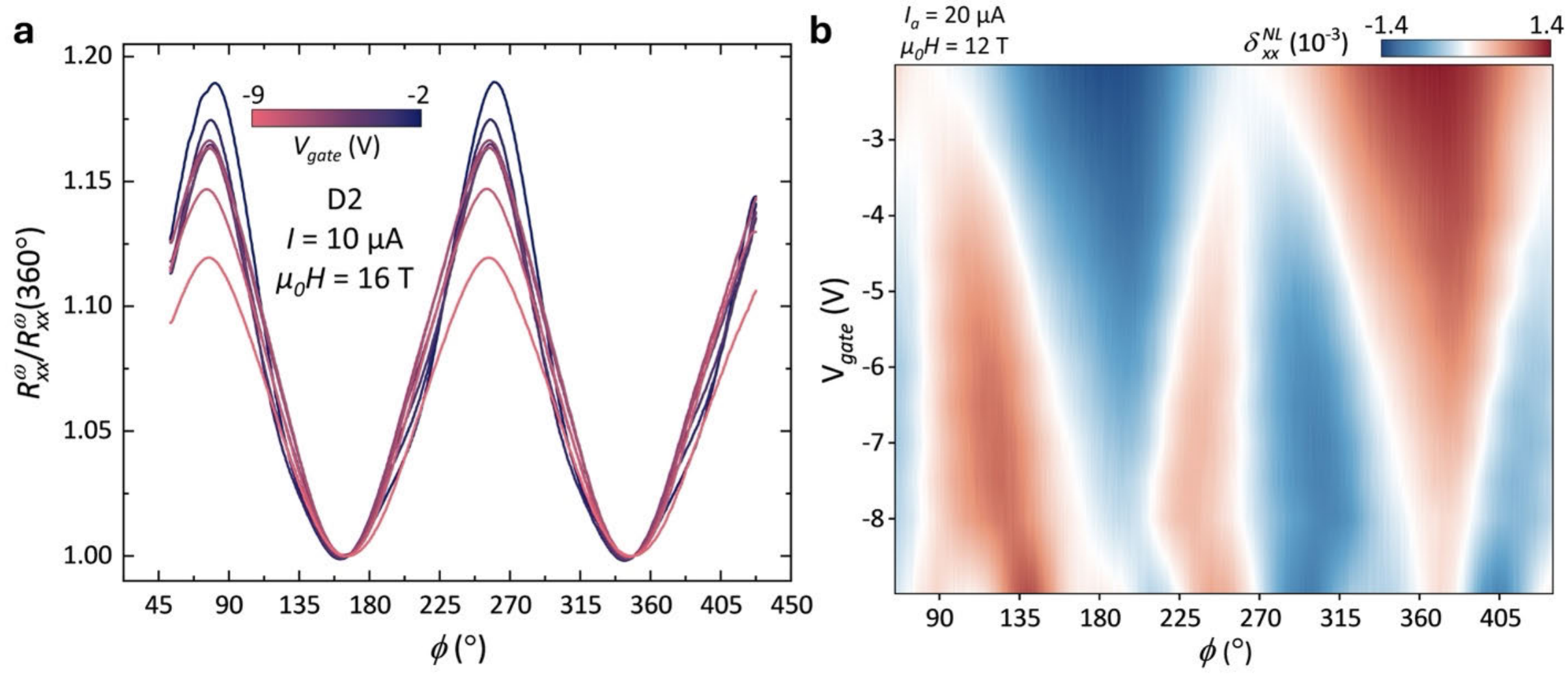

Extended Data Fig. 7 | **a**, angular dependence of $R^{\omega}_{xx}$ under different gate voltages. Unlike $R^{\omega}_{zz}$, the angular shift in $R^{\omega}_{xx}$ approximately remain fixed at ~10° within this gate voltage range. **b**, gate voltage mapping of $\delta^{NL}_{xx}$ for $I$ = 20 uA and $\mu_0 H$ = 12 T, illustrating the transition of $\delta^{NL}_{xx}$ from weak SOC to strong SOC regions, consistently capturing the previously observed features.

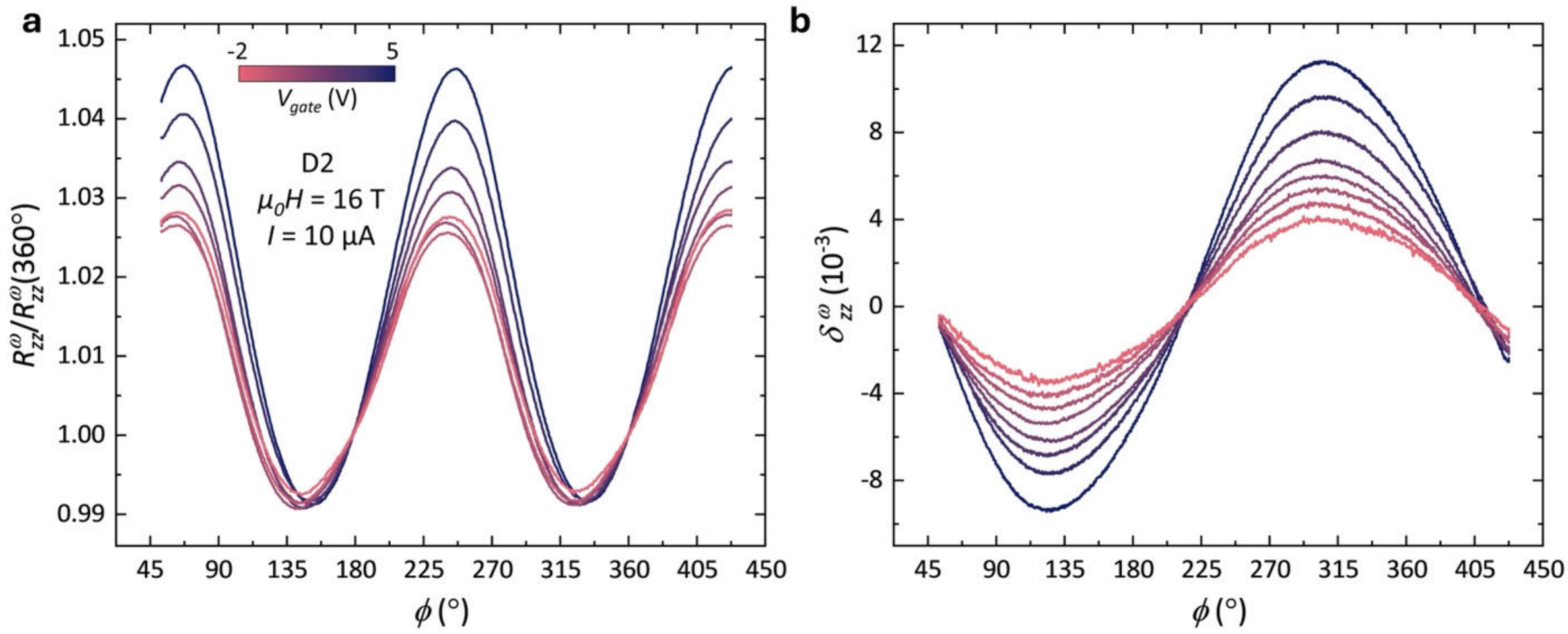

Extended Data Fig. 8 | **a**, **b** angular dependence of $R_{zz}^{\omega}$ (**a**) and $\delta_{zz}^{NL}$ (**b**) under extended gate voltage range (-2 V to 5 V) measured on D2. The angular shift of $R_{zz}^{\omega}$ and $\delta_{zz}^{NL}$ remains fixed, while the amplitude of $\delta_{zz}^{NL}$ continues to increase as SOC weakens, highlighting the significance of orbital effects.

# Supplementary Information for "Tunable interplay of orbital and spin magnetization in trigonal tellurium"

Zhenqi Hua [1,2,*], Chang Niu[3,4,*], Sandeep Joy[1,2,*], Pukun Tan[3,4], Gang Shi[1], Haoyang Liu[1], Jiaxing Guo[1], David Graf[2], Peide Ye[3,4], Cyprian Lewandowski[1,2,†], and Peng Xiong[1,†]

[1]*Department of Physics, Florida State University, Tallahassee, FL 32306*

[2]*National High Magnetic Field Laboratory, Tallahassee, FL 32310*

[3]*Elmore Family School of Electrical and Computer Engineering, Purdue University, West Lafayette, IN 47907*

[4]*Birck Nanotechnology Center, Purdue University, West Lafayette, IN 47907*

## S1. REVIEW OF LINEAR AND NON-LINEAR TRANSPORT

In this section, we outline the main steps in deriving the conductivity tensor in both the linear (electric field-independent) and nonlinear (linear in electric field) regimes. While we aim to keep the discussion as general as possible, we conclude with details specific to tellurium (Te). The derivation of the linear conductivity tensor is standard textbook material, but we include it here for the sake of completeness (for e.g. See Ref [S1]). The standard semiclassical Boltzmann formalism begins with the equations of motion for charge carriers:

$$\dot{\boldsymbol{r}} = \frac{1}{\hbar}\frac{\partial \varepsilon}{\partial \boldsymbol{k}}, \tag{S1}$$

$$\hbar\dot{\boldsymbol{k}} = -e\boldsymbol{E}, \tag{S2}$$

where $\varepsilon$ is the energy dispersion of the carriers. Under the assumptions of spatial homogeneity and the relaxation time approximation ($\tau$), the Boltzmann equation takes the form:

$$\frac{-e\boldsymbol{E}}{\hbar}\cdot\frac{\partial f}{\partial \boldsymbol{k}} = -\frac{f - f^0}{\tau}, \tag{S3}$$

where $f$ is the perturbed distribution function and $f^0$ is the equilibrium distribution function. Knowledge of $f$ allows us to determine the system's response, i.e., its conductivity. We assume that the solution for $f$ can be expanded as a power series in $\boldsymbol{E}$:

$$f = f^0 + f^1 + f^2 + \ldots \tag{S4}$$

Substituting the above expansion into the Boltzmann equation, the first- and second-order corrections can be obtained in terms of the equilibrium distribution:

$$f^1 = \frac{eE_\mu\tau}{\hbar}\frac{\partial f^0}{\partial k_\mu}, \tag{S5}$$

$$f^2 = \frac{e^2\tau^2 E_\mu E_\nu}{\hbar^2}\frac{\partial^2 f^0}{\partial k_\mu \partial k_\nu}. \tag{S6}$$

These expressions can be used to compute the linear and nonlinear conductivity tensors. The current from a single band is given by integrating the carrier velocity weighted by the distribution function over momentum space:

$$j_\alpha = \frac{1}{(2\pi)^2}\int d\boldsymbol{k}\,(-ev_\alpha)f. \tag{S7}$$

Substituting the series expansion of $f$, we find the linear conductivity tensor:

$$\sigma^0_{\alpha\beta} = \frac{e^2\tau}{(2\pi)^2\hbar}\int d\boldsymbol{k}\left(\frac{\partial\varepsilon}{\partial k_\alpha}\right)\left(\frac{\partial\varepsilon}{\partial k_\beta}\right)\left(-\frac{\partial f^0}{\partial\varepsilon}\right), \tag{S8}$$

and the second-order (nonlinear) conductivity tensor:

$$\sigma^1_{\alpha\beta\gamma} = \frac{e^3\tau^2}{(2\pi)^2\hbar^2}\int d\boldsymbol{k}\, v_\alpha\left(-\frac{\partial^2 f^0}{\partial k_\beta \partial k_\gamma}\right). \tag{S9}$$

Additionally, other contributions to nonlinear conductivity—such as those arising from the Berry curvature dipole, quantum geometry, and skew scattering—are not discussed here [S2]. However, our focus is specifically on the dependence of the (linear) in-plane magnetic field and its orientation on the longitudinal linear and nonlinear response. We should also like to point out that, the results discussed in the main text are carried out for constant chemical potential for all the models. We have verified however that the same trends would be seen if the simulations were plotted for a constant charge density.

In transport experiments, a fixed current is applied and the resulting voltage drop is measured. Consequently, what is directly extracted is the resistivity tensor (or resistance tensor), rather than the conductivity tensor. However, theoretical calculations, such as those based on the Boltzmann formalism, typically yield the conductivity tensor. Below, we present a systematic procedure for inverting the conductivity tensor order by order to obtain the corresponding resistivity tensor. The relation between current density and electric field, expressed in terms of the conductivity and resistivity tensors, is given by:

$$j_\alpha = \sigma^0_{\alpha\beta}E_\beta + \sigma^1_{\alpha\beta\gamma}E_\beta E_\gamma, \tag{S10}$$

and

$$E_\alpha = \rho^0_{\alpha\beta} j_\beta + \rho^1_{\alpha\beta\gamma} j_\beta j_\gamma. \tag{S11}$$

We substitute the components of the electric field $\boldsymbol{E}$ in terms of the current density $\boldsymbol{j}$ and the resistivity tensor into the expression for $j_\alpha$:

$$\begin{aligned} j_\alpha &= \sigma^0_{\alpha\beta} E_\beta + \sigma^1_{\alpha\beta\gamma} E_\beta E_\gamma \\ &= \sigma^0_{\alpha\beta}\left(\rho^0_{\beta\mu} j_\mu + \rho^1_{\beta\mu\nu} j_\mu j_\nu\right) + \sigma^1_{\alpha\beta\gamma}\left(\rho^0_{\beta\mu} j_\mu + \rho^1_{\beta\mu\nu} j_\mu j_\nu\right)\left(\rho^0_{\gamma\lambda} j_\lambda + \rho^1_{\gamma\lambda\eta} j_\lambda j_\eta\right) \\ &= \sigma^0_{\alpha\beta}\rho^0_{\beta\mu} j_\mu + \sigma^0_{\alpha\beta}\rho^1_{\beta\mu\nu} j_\mu j_\nu + \sigma^1_{\alpha\beta\gamma}\rho^0_{\beta\mu}\rho^0_{\gamma\lambda} j_\mu j_\lambda + O\left(j^3\right) \end{aligned} \tag{S12}$$

Solving the term linear in $\boldsymbol{j}$ reproduces the standard result from linear response theory:

$$\sigma^0_{\alpha\beta}\rho^0_{\beta\mu} = \delta_{\alpha\mu}, \tag{S13}$$

i.e., the linear resistivity tensor is the inverse of the linear conductivity tensor. At second order, we identify the resistivity tensor $\rho^1_{\alpha\beta\gamma}$ in terms of the conductivity tensor $\sigma^1_{\alpha\beta\gamma}$ as:

$$\rho^1_{\alpha\beta\gamma} = -\rho^0_{\alpha\mu}\sigma^1_{\mu\nu\lambda}\rho^0_{\nu\beta}\rho^0_{\lambda\gamma}. \tag{S14}$$

It is interesting to note that knowledge of the linear resistivity is required in order to invert the nonlinear conductivity tensor and obtain the nonlinear resistivity tensor.

## S2. REVIEW OF THEORETICAL MODELS OF TE

In this section, we summarize various theoretical models of Te that have been proposed in the literature. Te has been extensively studied as an elemental semiconductor since the 1970s, owing to its unique chiral crystal structure, band dispersion, spin texture, and particularly its pronounced photogalvanic effects [S3–S9]. The models we discuss in this section include: (a) a phenomenological model capturing the dispersion and associated spin texture, (b) a two-band model—considered either with two copies to reproduce the two valleys in Te or as a single-valley model with a radial spin texture, and (c) a six-band model (again with two copies for the two valleys) that incorporates orbital magnetization. Due to the inherent $C_3$ rotational symmetry of these models, as expected, we found that none of these models successfully reproduce the experimentally observed trends in their present form using the Boltzmann transport framework.

### S2.A. Phenomenological model

The bare band dispersion of the valence band in Te is given by [S10]

$$\varepsilon_v(\boldsymbol{k}) = -\frac{\hbar^2 k_x^2}{2m} - \frac{\hbar^2 k_z^2}{2m} + \sqrt{\alpha_z^2 k_z^2 + \Delta^2}. \tag{S15}$$

We introduce natural units of momentum and energy defined by the characteristic scales of the bare band structure discussed above:

$$k_0 \equiv \frac{m\alpha_z}{\hbar^2}, \quad \text{and} \quad \varepsilon_0 \equiv \frac{\hbar^2 k_0^2}{2m} = \frac{m\alpha_z^2}{2\hbar^2} = \frac{\alpha_z k_0}{2}. \tag{S16}$$

In terms of these rescaled units, the dispersion becomes

$$\varepsilon \equiv \frac{\varepsilon_v(k_0\boldsymbol{q})}{\varepsilon_0} = -q_x^2 - q_z^2 + \sqrt{4q_z^2 + d^2}, \tag{S17}$$

where $\boldsymbol{q} = \boldsymbol{k}/k_0$ and $d = \Delta/\varepsilon_0$. The numerical values for the energy and momentum scales are:

$$\varepsilon_0 = 3.95 \times 10^{-2}\,\text{eV}, \quad k_0 = 3.29 \times 10^{-2}\,\text{Å}^{-1}. \tag{S18}$$

The "radial" spin texture of Te introduces a Zeeman coupling to the bare dispersion when an in-plane magnetic field is applied, $\boldsymbol{B} = B_{\parallel}(\sin\phi, 0, \cos\phi)$ [S10]. The spin texture is given by:

$$\boldsymbol{s} \sim \left(0, 0, \frac{3k_z}{2\sqrt{\alpha_z^2 k_z^2 + \Delta^2}}\right). \tag{S19}$$

The spin texture couples to the magnetic field, contributing the following Zeeman term to the energy dispersion:

$$\begin{aligned} \delta\varepsilon &\sim \frac{\mu_0 B_{\parallel}}{\varepsilon_0} s_z \cos\phi \\ &\sim \left(\frac{\mu_0 B_{\parallel}}{\varepsilon_0}\right) \frac{3q_z}{\sqrt{4q_z^2 + d^2}} \cos\phi. \end{aligned} \tag{S20}$$

The corresponding dimensionless parameter and magnetic energy scale are:

$$d \equiv \frac{\Delta}{\varepsilon_0} = 1.51, \quad B_0 \equiv \left(\frac{\mu_0 B_{\parallel}}{\varepsilon_0}\right) \approx 1.4 \times 10^{-3}, \mathrm{eV/T}. \tag{S21}$$

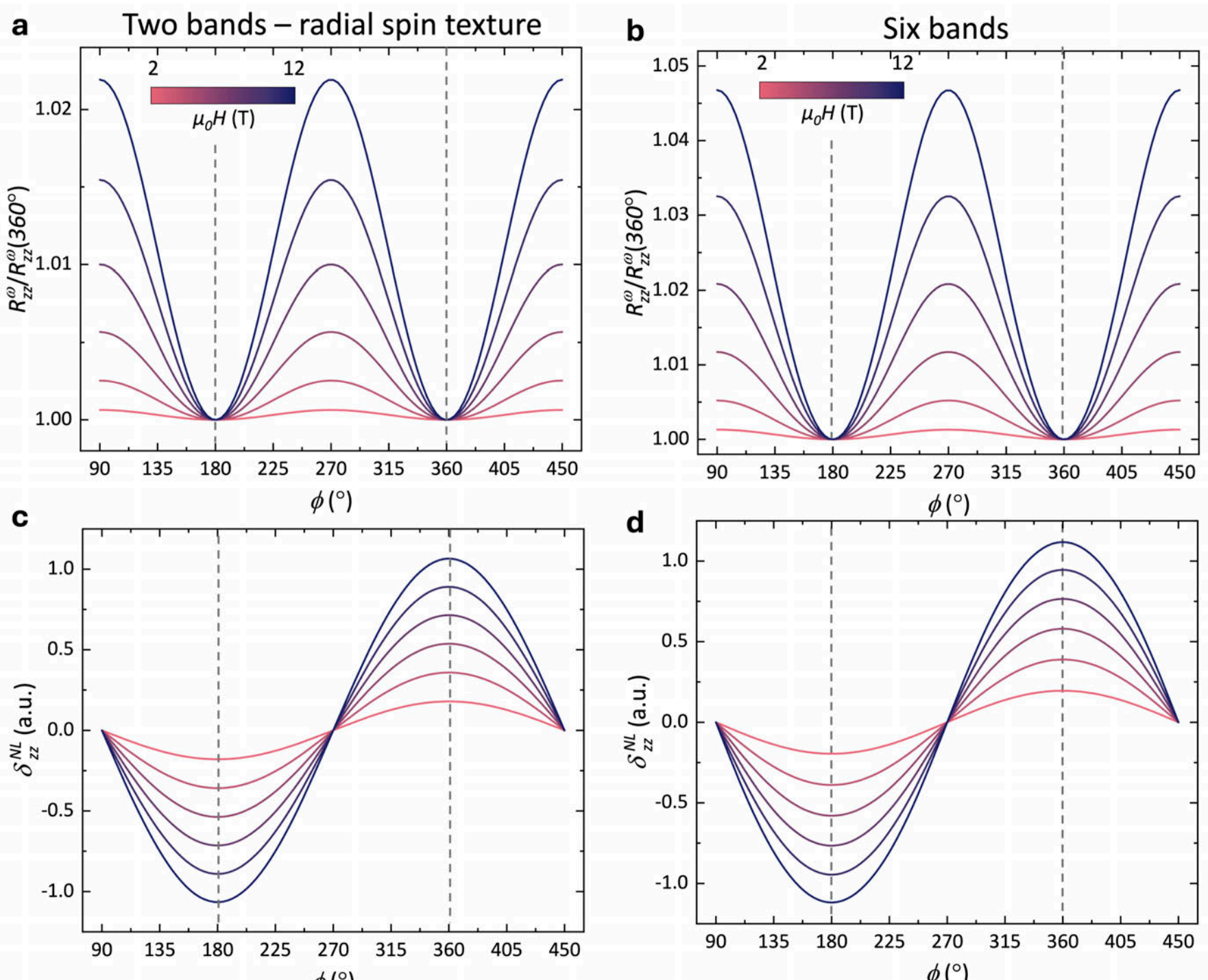

FIG. S1. These figures show the linear and non-linear resistivity as a function of the direction of the in-plane magnetic field, calculated using the Boltzmann semiclassical transport formalism. Panels (a) and (c) correspond to an energy dispersion that includes Zeeman coupling to the radial spin texture in a single-valley model [S11]. As in Fig. 4(a) and (d) of the main text (which considers two valleys with identical $s_z$ and opposite $s_x$), varying the magnetic field strength does not shift the oscillation minima associated with the collinear and anti-collinear directions. Panels (b) and (d) show results from the six-band model in Eq. S24, which includes two valleys and accounts for both spin and orbital magnetization. Even in this case, the positions of the collinear and anti-collinear minima remain unchanged with increasing magnetic field.

### S2.B. Two-band model of Te

In the Te literature, a low-energy two-band model is frequently used to describe the valence band [S12, S13]. (Close to the Fermi level, there are six bands in total; we will discuss a six-band model below.) The two-band model, written in the spin basis, is given by:

$$\frac{\hat{H}}{\varepsilon_0} = -\left(q_x^2 + q_z^2\right)\mathbb{I} + 2q_z\sigma_z + d\sigma_x. \tag{S22}$$

Here, $\sigma_x$ and $\sigma_z$ are Pauli matrices. The above given two-band model reproduces the dispersion relation in Eq. S17. Since the Hamiltonian is expressed in the spin basis, one can extract the spin texture of the valence band as follows:

$$s_z = \frac{2q_z}{\sqrt{4q_z^2 + d^2}} \quad \text{and} \quad s_x = \frac{d}{\sqrt{4q_z^2 + d}}. \tag{S23}$$

The two-band model breaks time-reversal symmetry: under time reversal, $s_x$ does not map to $-s_x$, which is not surprising, as the $H$ point around which this low-energy Hamiltonian is expanded is not a time-reversal-invariant momentum point [S12, S14]. In fact, there exists another high-symmetry point $H'$, which is related to $H$ by time reversal and restores time-reversal symmetry. The relation between two valleys implies that $s_x(H) = -s_x(H')$, suggesting that $d$ must change sign between them. A related single-valley version of the model that preserves time-reversal symmetry has been studied in the literature by replacing $d$ with $\text{sign}(q_x), d$, leading to a radial spin texture [S11]. However, even this variant fails to reproduce the experimentally observed shift in the resistivity minima and maxima (see Fig S1).

### S2.C. Six-band model of Te

The most complicated model we considered is a six-band model, which includes four valence bands and two conduction bands [S12]. Here we reproduce the six-band Hamiltonian:

$$h_{\boldsymbol{k}} = \begin{pmatrix} E_0 + \beta_c q_z + \lambda_c & \delta_{1c} q_+^2 + A q_- & 0 & P_1 q_+ & P_3 q_z & P_2 q_- \\ \delta_{1c} q_-^2 + A^* q_+ & E_0 - \beta_c q_z + \lambda_c & P_1 q_- & 0 & P_2 q_+ & -P_3 q_z \\ 0 & P_1 q_+ & \beta_v q_z + \lambda_v & d_2 & 0 & P_4 q_- \\ P_1 q_- & 0 & d_2 & -\beta_v q_z + \lambda_v & P_4 q_+ & 0 \\ P_3 q_z & P_2 q_- & 0 & P_4 q_- & \beta_h q_z - 2d_1 + \lambda_h & \delta_{1h} q_+^2 + B q_- \\ P_2 q_+ & -P_3 q_z & P_4 q_+ & 0 & \delta_{1h} q_-^2 + B^* q_+ & -\beta_h q_z - 2d_1 + \lambda_h \end{pmatrix}. \tag{S24}$$

The symbols and parameters in Eq. S24 are given in Table I and Table II.

| Symbol | Expression |
|---|---|
| $A$ | $\alpha_c + i\delta_{2c} q_z$ |
| $B$ | $\alpha_h + i\delta_{2h} q_z$ |
| $q_\pm$ | $q_x \pm i q_y$ |
| $q_\perp^2$ | $q_x^2 + q_y^2$ |
| $\lambda_i$ | $\frac{\hbar^2}{2m_\parallel^i} q_z^2 + \frac{\hbar^2}{2m_\perp^i} q_\perp^2$ |

TABLE I. List of symbols used in the Hamiltonian in Eq. S24 and their expressions.

The six-band model reduces to the two-band Hamiltonian (Eq. S22) upon projection onto the top two valence bands and exhibits an analogous spin texture. Unlike the previous models however it has non-zero orbital magnetization, which is plotted in Fig S2 along with their spin-magnetization. Specifically, we find that the model has a non-zero $m_z$ component of orbital magnetization that is odd under momentum $q_z$. The $m_x$, however has not such symmetry and thus does not reproduce the shift from the collinear and anti-collinear directions (see Fig S1) - see also the discussion of the next section.

## S3. SYMMETRY ANALYSIS OF NON-LINEAR RESISTIVITY: WHAT COULD GIVE A SHIFT IN THE MINIMA?

Here, we aim to identify the minimal ingredients necessary to account for the experimentally observed trends in the positions of resistivity minima and maxima. In particular, we motivate the need for an orbital magnetization along the $x$-direction that is odd in $q_z$, a feature that naturally emerges in the intrinsic $C_3$ symmetry-breaking helical chain model introduced later.

| Parameter | Value | Parameter | Value |
|---|---|---|---|
| $\beta_v$ | $1.99\,\varepsilon_0 \cdot k_0^{-1}$ | $E_0$ | $10\,\varepsilon_0$ |
| $\beta_c$ | $-0.63\,\varepsilon_0 \cdot k_0^{-1}$ | $\alpha_c$ | $0.055\,\varepsilon_0 \cdot k_0^{-1}$ |
| $\beta_h$ | $0.22\,\varepsilon_0 \cdot k_0^{-1}$ | $\alpha_h$ | $0.039\,\varepsilon_0 \cdot k_0^{-1}$ |
| $P_1$ | $2.79\,\varepsilon_0 \cdot k_0^{-1}$ | $\delta_{1c}$ | $0.010\,\varepsilon_0 \cdot k_0^{-2}$ |
| $P_2$ | $2.95\,\varepsilon_0 \cdot k_0^{-1}$ | $\delta_{2c}$ | $0.062\,\varepsilon_0 \cdot k_0^{-2}$ |
| $P_3$ | $1.66\,\varepsilon_0 \cdot k_0^{-1}$ | $\delta_{1h}$ | $0.003\,\varepsilon_0 \cdot k_0^{-2}$ |
| $P_4$ | $1.61\,\varepsilon_0 \cdot k_0^{-1}$ | $\delta_{2h}$ | $0.024\,\varepsilon_0 \cdot k_0^{-2}$ |
| $\frac{\hbar^2}{2m_{\parallel}^{v}}$ | $-0.99\,\varepsilon_0 \cdot k_0^{-2}$ | $\frac{\hbar^2}{2m_{\perp}^{v}}$ | $-0.89\,\varepsilon_0 \cdot k_0^{-2}$ |
| $\frac{\hbar^2}{2m_{\parallel}^{c}}$ | $1.45\,\varepsilon_0 \cdot k_0^{-2}$ | $\frac{\hbar^2}{2m_{\perp}^{c}}$ | $1.57\,\varepsilon_0 \cdot k_0^{-2}$ |
| $\frac{\hbar^2}{2m_{\parallel}^{h}}$ | $-0.93\,\varepsilon_0 \cdot k_0^{-2}$ | $\frac{\hbar^2}{2m_{\perp}^{h}}$ | $-1.26\,\varepsilon_0 \cdot k_0^{-2}$ |
| $d_1$ | $1.34\,\varepsilon_0$ | $d_2$ | $1.51\,\varepsilon_0$ |

TABLE II. List of parameters used in the Hamiltonian in Eq. S24 and their values. As defined previously, $\varepsilon_0 = 3.95 \times 10^{-2}$ eV, $k_0 = 3.29 \times 10^{-2}$ Å$^{-1}$

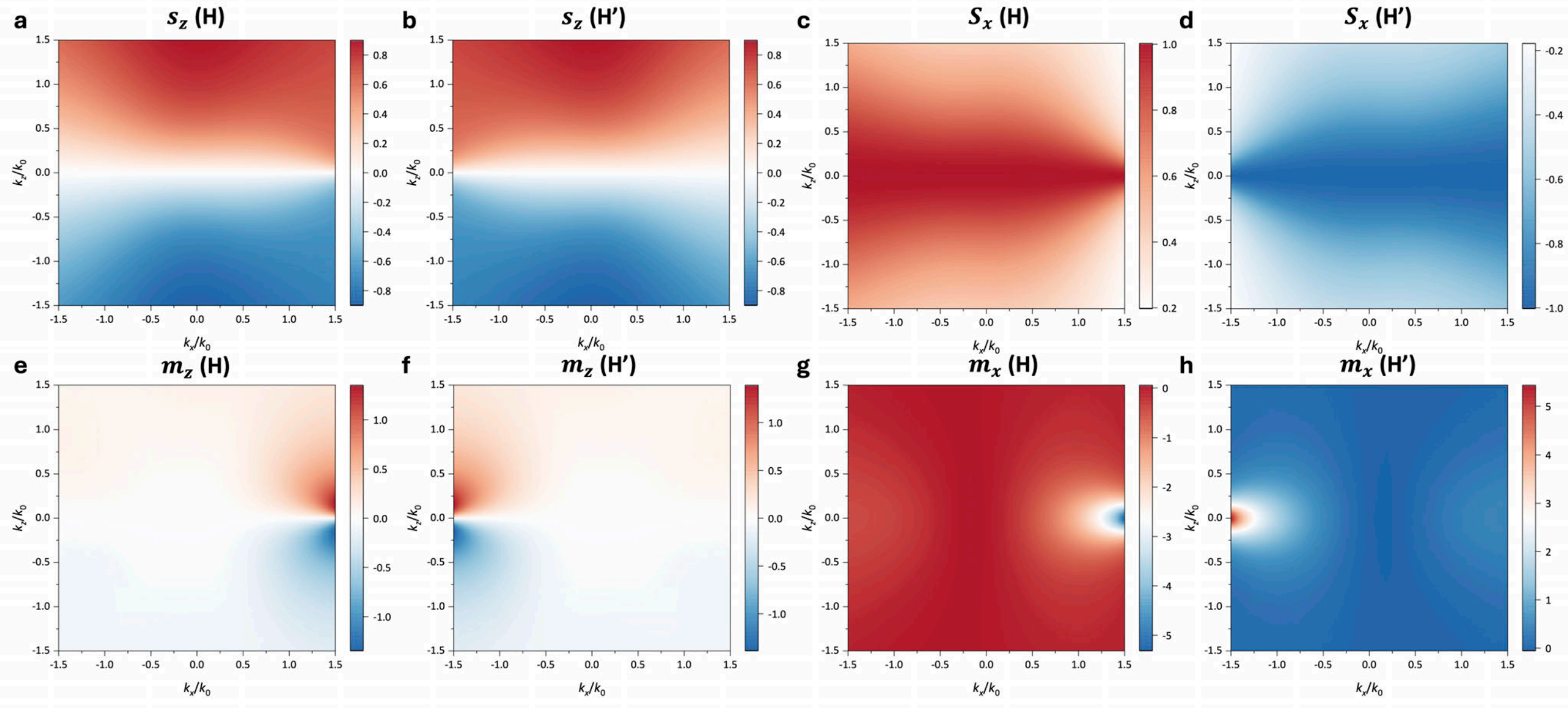

FIG. S2. The orbital $m_{x/z}$ and spin $s_{x/z}$ magnetization calculated from the Hamiltonian in Eq. S24 is plotted. Title subscript in figure title denotes valley $H/H'$, which are related to each other by $d_2 \to -d_2$.

The dispersion of the valence band in the presence of a magnetic field includes a Zeeman coupling term, which gives rise to magnetotransport. Let us examine this more carefully. The magnetic field dependent part of the dispersion takes the form (remember $B_0 \equiv (\mu_0 B_{\parallel}/\varepsilon_0)$):

$$\delta\varepsilon = -B_0 \cos\phi(s_z + m_z) - B_0 \sin\phi(s_x + m_x)\,, \tag{S25}$$

which can be recast as:

$$\delta\varepsilon = -B_0 \sqrt{(s_z + m_z)^2 + (s_x + m_x)^2} \cos(\phi - \phi_0), \tag{S26}$$

where

$$\phi_0 = \arctan\left[\frac{s_x + m_x}{s_z + m_z}\right]. \tag{S27}$$

To ensure that the linear magnetoresistance $\rho_{zz}$ is 180°-periodic (mandated by the time-reversal symmetry of the equilibrium state since Te is non-magnetic), the vector $\boldsymbol{m}$ must be odd under $\boldsymbol{q}$, thereby preserving time-reversal symmetry. Given the helical chain structure of Te, a natural candidate for $\boldsymbol{m}$ is the orbital magnetization of the Bloch states, arising from the solenoidal character of the system. The above Eq. (S27) also shows that unless the system has a non-zero spin or orbital magnetization along the $x$ direction, the maxima/minima of the dispersion are locked to the collinear/anti-collinear directions (i.e.$\phi_0 = 0, 180°$). One could also imagine a magnetization term along the $x$-direction that is odd under $q_x$. However, as shown in Fig. 4(b) and (e) of the main text, such a term does not shift the minima or maxima away from the collinear and anti-collinear directions. This is consistent with the result in Fig. S1(a) and (c), where an $s_x$ component that is odd under $q_x$ similarly fails to produce any shift.

## S4. ONE-DIMENSIONAL TE AS A HELICAL CHAIN: A SIMPLIFIED MODEL

In this section, we discuss the modern theory of orbital magnetization in the context of Te, modeled as a chiral helical chain. We begin with a review of how to calculate orbital magnetization in one-dimensional systems. We then construct a minimal model based on one-dimensional Te chains with explicitly broken $C_3$ symmetry and show that it gives rise to orbital magnetization perpendicular to the helical axis, which explains the experimentally observed results. We note in passing that the condensed matter analogue of (quantum) solenoid character in Te has been previously recognized in the literature [S10, S15, S16]. We also want to emphasize that the proposed minimal model of Te is phenomenological in nature, with its sole purpose of demonstrating how $C_3$ symmetry breaking allows the system to exhibit transverse orbital magnetization. Future work would be necessary to identify the microscopic origins of the $C_3$ symmetry breaking, which could demonstrate that the effective model used here is not appropriate.

### S4.A. Modern theory of orbital magnetization in one dimension

Here, we provide a short summary of the modern theory of orbital magnetization that is used to compute magnetization of the helical chain. The notion that a finitely spread wavepacket can exhibit self-rotation around its center of mass underlies the modern theory of orbital magnetization (see Ref [S17], and references therein). The orbital magnetization behaves analogously to spin magnetization, coupling to external magnetic fields through a Zeeman-like interaction. In two-dimensional materials, out-of-plane orbital magnetization has sparked considerable interest over the past few decades. From a classical perspective solenoid, it is evident that a charged particle undergoing one-dimensional helical motion acquires an orbital magnetic moment along its direction of motion. We will demonstrate how this intuition can be placed on firm theoretical footing. The orbital magnetization for a band labeled by index $n$ at momentum $\boldsymbol{q}$ is given by[S17–S19]

$$m_n^a(\boldsymbol{q}) = \frac{ie}{2\hbar}\sum_{m\neq n}\epsilon^{abc}\frac{\langle u_n(\boldsymbol{q})|\, v^b\, |u_m(\boldsymbol{q})\rangle\langle u_m(\boldsymbol{q})|\, v^c\, |u_n(\boldsymbol{q})\rangle}{\varepsilon_n(\boldsymbol{q})-\varepsilon_m(\boldsymbol{q})} \tag{S28}$$

where $v^b$ is the velocity operator in the $b$-direction, $|u_n(\boldsymbol{q})\rangle$ is the Bloch wavefunction of the $n^{\text{th}}$ band at momentum $\boldsymbol{q}$, and $\varepsilon_n(\boldsymbol{q})$ is the energy of the $n^{\text{th}}$ band at momentum $\boldsymbol{q}$. In three-dimensional materials, this is typically given by the operator $\partial H/\partial q_b$. In two-dimensional systems lying in the $x$-$y$ plane, this expression implies that the only nonzero component of orbital magnetization is along the $z$-direction. In strictly one-dimensional systems, the expression predicts vanishing orbital magnetization in all directions.

However, care must be taken when defining the velocity operator if any spatial direction is finite, such as in a thin slab of a two-dimensional system or a one-dimensional system with internal structure within the unit cell. The ambiguity can be addressed by defining the velocity operator via the Heisenberg equation of motion:

$$v^a = \frac{-i}{\hbar}\left[x^a, H\right]. \tag{S29}$$

Plugging this into the matrix element of the velocity operator yields

$$\begin{aligned}\langle u_n(\boldsymbol{q})|\, v^a\, |u_m(\boldsymbol{q})\rangle &= \langle u_n(\boldsymbol{q})|\,\frac{-i}{\hbar}\left[x^a, H\right]|u_m(\boldsymbol{q})\rangle\\ &= \frac{i}{\hbar}\left(\varepsilon_n(\boldsymbol{q})-\varepsilon_m(\boldsymbol{q})\right)\langle u_n(\boldsymbol{q})|\, x^a\, |u_m(\boldsymbol{q})\rangle\end{aligned} \tag{S30}$$

The above expression makes it clear that the position operator matrix element in the Bloch basis is ill-defined when the corresponding spatial direction is periodic or infinite. However, in systems that are finite along some directions, this formulation becomes meaningful and can be used to compute the orbital magnetization properly [S19–S23].

To evaluate the orbital magnetization in a one-dimensional chain numerically later, the Bloch wavefunction can be decomposed in the Fourier basis as follows:

$$\begin{aligned} |u_n(\boldsymbol{q})\rangle &= \sum_{\alpha,\sigma} a^{\alpha}_{n,\sigma} \, |\boldsymbol{q}^{\alpha}, \sigma\rangle \\ &= \frac{1}{\sqrt{N}} \sum_{\alpha,\sigma,j} a^{\alpha}_{n,\sigma} \exp\left[-i\boldsymbol{q}\cdot\boldsymbol{r}^{\alpha}_{j}\right] |j^{\alpha}, \sigma\rangle \,, \end{aligned} \tag{S31}$$

where $\sigma$ is the spin index, $\alpha$ labels the atomic position within the unit cell, $j$ labels the unit cell in real space, and $n$ denotes the band index. The coefficient $a^{\alpha}_{n,\sigma}$ is the component of the eigenvectors from diagonalizing the Hamiltonian in the momentum space. The matrix element of the position operator between two Bloch states is then given by:

$$\begin{aligned} \langle u_n(\boldsymbol{q})| \, x^a \, |u_m(\boldsymbol{q})\rangle &= \frac{1}{N} \sum_{\alpha,\beta,j,l,\sigma,\sigma'} a^{\alpha}_{n,\sigma} a^{\beta\star}_{m,\sigma'} \exp\left[-i\boldsymbol{q}.\boldsymbol{r}^{\alpha}_{j}\right] \exp\left[i\boldsymbol{q}.\boldsymbol{r}^{\beta}_{l}\right] \left\langle l^{\beta}, \sigma' \right| x^a \left| j^{\alpha}, \sigma \right\rangle \\ &= \frac{1}{N} \sum_{\alpha,\beta,j,l,\sigma,\sigma'} a^{\alpha}_{n,\sigma} a^{\beta\star}_{m,\sigma'} \exp\left[-i\boldsymbol{q}.\boldsymbol{r}^{\alpha}_{j}\right] \exp\left[i\boldsymbol{q}.\boldsymbol{r}^{\beta}_{l}\right] \boldsymbol{r}^{a} \delta_{jl} \delta_{\alpha\beta} \delta_{\sigma,\sigma'} \\ &= \sum_{\alpha} a^{\alpha}_{n,\sigma} a^{\alpha\star}_{m,\sigma} \boldsymbol{r}^{a} \end{aligned} \tag{S32}$$

Here, $\boldsymbol{r}^a$ denotes the $a$-th component of the position of atom $\alpha$ within the unit cell.

### S4.B. Simplified one-dimensional model

Given that the symmetry analysis in the previous section suggests the presence of an orbital magnetization along the $x$-direction that is odd in momentum along the (helical) $z$-direction—potentially responsible for the observed trends in both linear and non-linear transport—we now introduce a minimal model of a one-dimensional Te chain (Te is known to consist of stacked one-dimensional helical chains.).

We consider a minimal model of a helical chain with three atoms per unit cell, each hosting a single $s$ orbital. While a more realistic model would include four valence orbitals per site (one $s$ and three $p$ orbitals) and account for the orientation of the $p$ orbitals in determining tunneling elements, our simplified model suffices to capture the essential features.

$$\begin{aligned} \boldsymbol{r_1} &= \begin{pmatrix} a & 0 & 0 \end{pmatrix} \\ \boldsymbol{r_2} &= \begin{pmatrix} -\frac{a}{2} & \frac{\sqrt{3}a}{2} & \frac{c}{3} \end{pmatrix} \\ \boldsymbol{r_3} &= \begin{pmatrix} -\frac{a}{2} & -\frac{\sqrt{3}a}{2} & \frac{2c}{3} \end{pmatrix} \end{aligned} \tag{S33}$$

The basis lattice vector is defined as:

$$\boldsymbol{r}^{\alpha}_{j} = \boldsymbol{R}_j + \boldsymbol{r}_{\alpha}, \tag{S34}$$

where

$$\boldsymbol{R}_j = j\boldsymbol{R} = j\begin{pmatrix} 0 & 0 & c \end{pmatrix} \tag{S35}$$

The hopping paths between atoms are given by $\boldsymbol{r}_{ij} = \boldsymbol{r}_i - \boldsymbol{r}_j$.

### S4.C. Minimal Tight-Binding Model – without spin-orbit coupling

The simplest version of the Hamiltonian is a tight-binding model with uniform tunneling between sites:

$$\hat{H} = -t \sum_{j=1}^{N} \left( c^{\dagger}_{j,1} c_{j,2} + c^{\dagger}_{j,2} c_{j,3} + c^{\dagger}_{j,3} c_{j+1,1} + h.c. \right), \tag{S36}$$

where $c_{j,\alpha}$ and $c^\dagger_{j,\alpha}$ are annihilation and creation operators at site $\alpha$ in unit cell $j$. Their momentum-space counterparts are defined as:

$$
\begin{aligned}
c_{j,\alpha} &= \frac{1}{\sqrt{N}} \sum_{\boldsymbol{q}} \exp\left[-i\boldsymbol{q}\cdot\boldsymbol{r}_j^\alpha\right] c_{\boldsymbol{q},\alpha} \\
c^\dagger_{j,\alpha} &= \frac{1}{\sqrt{N}} \sum_{\boldsymbol{q}} \exp\left[i\boldsymbol{q}\cdot\boldsymbol{r}_j^\alpha\right] c^\dagger_{\boldsymbol{q},\alpha}
\end{aligned}
\tag{S37}
$$

Transforming to momentum space, the Hamiltonian becomes:

$$
\hat{H} = -t \sum_{\boldsymbol{q}} \begin{pmatrix} c^\dagger_{\boldsymbol{q},1} & c^\dagger_{\boldsymbol{q},2} & c^\dagger_{\boldsymbol{q},3} \end{pmatrix} \begin{pmatrix} 0 & \exp\left[i\boldsymbol{q}\cdot\boldsymbol{r}_{21}\right] & \exp\left[-i\boldsymbol{q}\cdot\boldsymbol{r}_{13}\right] \\ \exp\left[-i\boldsymbol{q}\cdot\boldsymbol{r}_{21}\right] & 0 & \exp\left[i\boldsymbol{q}\cdot\boldsymbol{r}_{32}\right] \\ \exp\left[i\boldsymbol{q}\cdot\boldsymbol{r}_{13}\right] & \exp\left[-i\boldsymbol{q}\cdot\boldsymbol{r}_{32}\right] & 0 \end{pmatrix} \begin{pmatrix} c_{\boldsymbol{q},1} \\ c_{\boldsymbol{q},2} \\ c_{\boldsymbol{q},3} \end{pmatrix}
\tag{S38}
$$

For our 1D helical chain, we consider only momentum along $z$, i.e.,

$$
\boldsymbol{q} = \begin{pmatrix} 0 & 0 & q_z \end{pmatrix}.
\tag{S39}
$$

Obtaining a simplified Hamiltonian:

$$
\hat{H}_{q_z} = -t \begin{pmatrix} 0 & \exp\left[\frac{iq_z c}{3}\right] & \exp\left[-\frac{iq_z c}{3}\right] \\ \exp\left[-\frac{iq_z c}{3}\right] & 0 & \exp\left[\frac{iq_z c}{3}\right] \\ \exp\left[\frac{iq_z c}{3}\right] & \exp\left[-\frac{iq_z c}{3}\right] & 0 \end{pmatrix}
\tag{S40}
$$

The eigenvalues of this Hamiltonian can be obtained analytically:

$$
\varepsilon_\ell\left(q_z\right) = -2t\cos\left(\frac{cq_z}{3} + \frac{2\pi\ell}{3}\right),
\tag{S41}
$$

where $\ell = 0, \pm 1$ correspond to the three $C_3$ eigenvalues (pseudo-angular momentum). The associated eigenvectors are:

$$
|\ell\rangle = \frac{1}{\sqrt{3}} \begin{pmatrix} 1 \\ \exp\left(\ell\frac{i2\pi}{3}\right) \\ \exp\left(-\ell\frac{i2\pi}{3}\right) \end{pmatrix}
\tag{S42}
$$

While preparing this manuscript, we became aware of an independent study that analyzes the same Hamiltonian, but which focus is on transport along the z-direction [S24]. Unfortunately the model of Eq. (S40) is too minimal, as it preserves the $C_3$ symmetry between the sites, to account for the experimental observations, since it only has an orbital magnetization along $z$ direction, $\propto a^2\sin\left(q_z c/3\right)$. By adding an asymmetry between sites (by onsite energy or inequivalent tunneling terms), and hence break the $C_3$ symmetry, one could obtain the orbital magnetization along $x$ and $y$-direction that is odd in $q_z$ - a procedure we follow in the next subsection.

### S4.D. Minimal Tight-Binding Model – with spin-orbit coupling

A spin texture locked to the momentum along the helical axis is a distinctive feature of Te. Such a texture can be realized by incorporating spin-orbit coupling into the model. To achieve this, we need the initially spin-degenerate band to split such that spin-up states have energy proportional to $q_z$ and spin-down states to $-q_z$, which can be implemented by introducing a term proportional to $\pm\sin q_z$ for the two spin flavors. Concretely, we add the following term:

$$
\hat{H}_{\mathrm{SOC}} = t_I \sum_{j=1}^{N} \sum_{\sigma} \mathrm{sign}(\sigma)\left(ic^\dagger_{j,1,\sigma}c_{j,2,\sigma} + ic^\dagger_{j,2,\sigma}c_{j,3,\sigma} + ic^\dagger_{j,3,\sigma}c_{j+1,1,\sigma} + \text{h.c.}\right),
\tag{S43}
$$

where $\mathrm{sign}(\sigma)$ distinguishes the spin flavors. To reproduce the approximately constant spin polarization along the $s_x$ direction, we also include an on-site spin-flip term:

$$
\hat{H}_{\mathrm{SOC,\ onsite}} = \Delta \sum_{j,\alpha,\sigma\neq\sigma'} c^\dagger_{j,\alpha,\sigma}c_{j,\alpha,\sigma'}.
\tag{S44}
$$

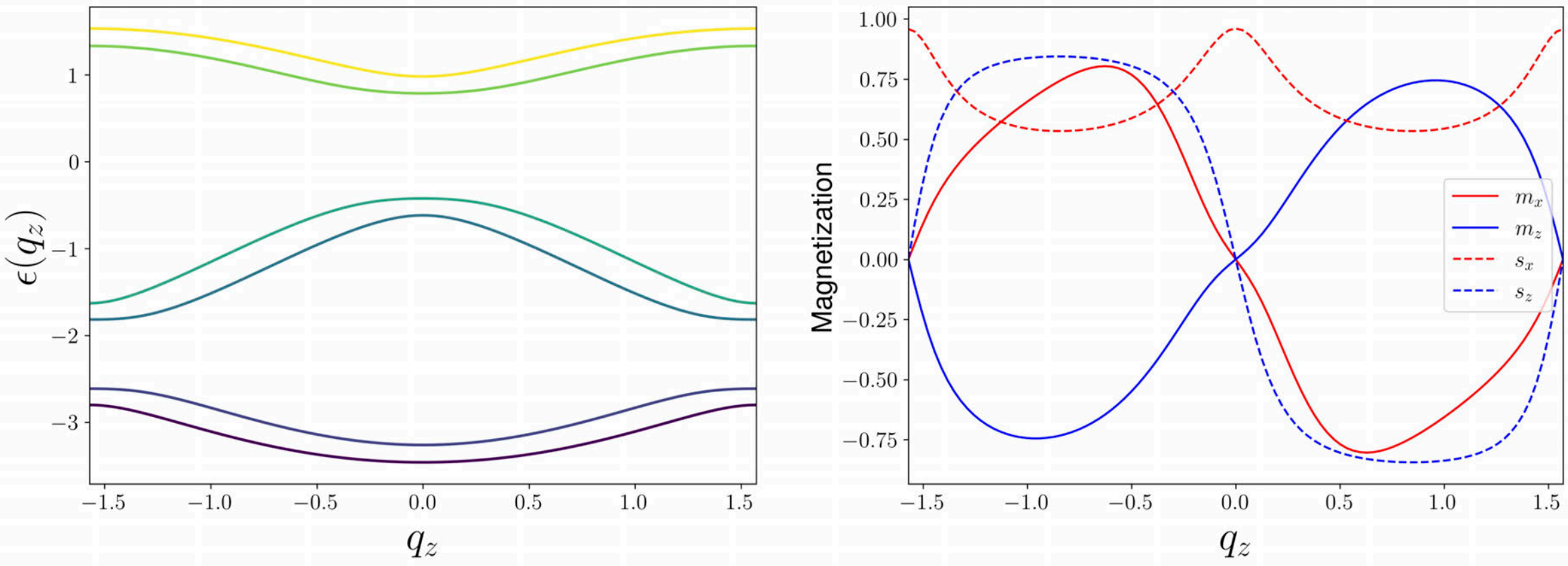

FIG. S3. (a) Band structure calculated from Eq. S45 using the parameters: $t_{12} = 1 + 0.1i$, $t_{23} = 1.1 + 0.1i$, $t_{31} = 1.2 + 0.1i$, $\Delta = 0.1$, $\nu_1 = -2$, $\nu_2 = -1$, $\nu_3 = 0$, and $c = 2$. For the fourth highest energy band, we show the spin and orbital components $s_{x/z}$ and $m_{x/z}$.

Finally, by allowing for site-dependent on-site energies $\nu_\alpha$, we arrive at the general form of the one-dimensional helical chain model. The following model is designed to faithfully capture both the spin and orbital magnetization along $q_z$ near the $H/H'$ points of three-dimensional Te:

$$\hat{H}(q_z) = \begin{bmatrix} \nu_1 & \Delta & -t_{12}\exp[i\Phi] & 0 & -t_{31}^\star \exp[-i\Phi] & 0 \\ \Delta & \nu_1 & 0 & -t_{12}^\star \exp[i\Phi] & 0 & -t_{31}\exp[-i\Phi] \\ -t_{12}^\star \exp[-i\Phi] & 0 & \nu_2 & \Delta & -t_{23}\exp[i\Phi] & 0 \\ 0 & -t_{12}\exp[-i\Phi] & \Delta & \nu_2 & 0 & -t_{23}^\star \exp[i\Phi] \\ -t_{31}\exp[i\Phi] & 0 & -t_{23}^\star \exp[-i\Phi] & 0 & \nu_3 & \Delta \\ 0 & -t_{31}^\star \exp[i\Phi] & 0 & -t_{23}\exp[-i\Phi] & \Delta & \nu_3 \end{bmatrix} . \tag{S45}$$

For brevity, we have defined $\Phi = q_z c/3$.

The current model has two major limitations. First, it is a simplified one-dimensional representation of a three-dimensional material, constructed using symmetry arguments and physical intuition. While it captures the magnetization trends reasonably well and qualitatively explains the experimental observations, a more rigorous ab-initio derivation is necessary for firm justification and quantitative comparison between theory and experiment. We leave this for future work.

### S4.E. Effect of surface Rashba SOC

It is necessary to emphasise that the symmetry-based analysis on the non-linear transport does not uniquely point towards orbital magnetization (and inherent $C_3$ symmetry breaking) as a mechanism. If a spin structure develops that has a non-collinear component with appropriate momentum ($x$ component being odd in $k_z$) is present, then the experimental trends could similarly be reproduced. In fact, such spin-structure in fact arises naturally in thin-films of Te, given that Te thin films break inversion symmetry at the interface with the substrate. With no inversion symmetry being present, Rashba spin–orbit coupling (SOC) is allowed and has the following structure

$$\hat{H}_{\text{Rashba}} = \alpha\, \boldsymbol{\sigma} \cdot (\hat{y} \times \boldsymbol{k}) , \tag{S46}$$

leading to the following spin texture:

$$\begin{aligned} s_x &\propto k_z \\ s_z &\propto k_x \end{aligned} \tag{S47}$$

Notably, as hinted above, this spin texture is consistent with the symmetry required to reproduce the experimentally observed shift of the resistivity minima. Motivated by this observation, we compute both the linear and nonlinear resistivities, including this interfacial Rashba SOC together with the bulk radial SOC. The resulting behavior is presented in Fig. S4.c These plots are consistent with experimental trends and the necessary Rashba coupling strength (~eVÅ is also consistent with the typical values seen in thin-films of Te. As argued in the main text, the Rashba SOC component however can be excluded due to thickness and doping dependence analysis, leaving orbital magnetization or intrinsic non-collinear spin structure of Te, which in itself would be highly unlikely, as the two possible origins of the experimental trends.

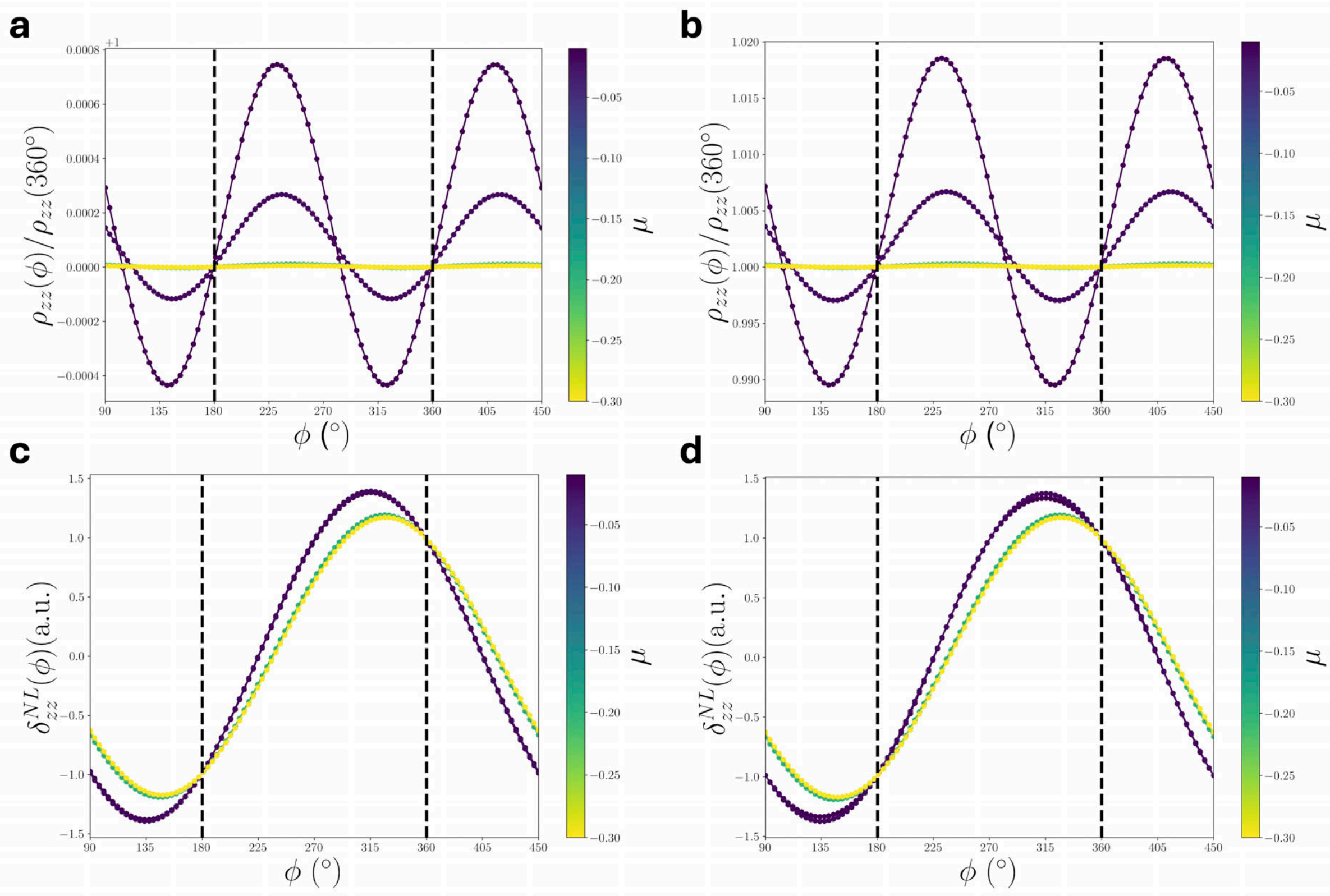

FIG. S4. (a), (b) Linear longitudinal resistivity as function of in plane field for $B = 2, 10\,T$. (c), (d) The corresponding $\delta_{NL}$ is shown. The Rashba coupling strength is taken as $\alpha = 0.3\,\mathrm{eV\AA}$ to match the experimentally observed shift.